\documentclass[useAMS,usenatbib]{mn2e}

\usepackage{txfonts}

\usepackage[T1]{fontenc}
\usepackage{ae,aecompl}

\usepackage{graphicx}	% Including figure files
\usepackage{wasysym}
\usepackage{morefloats}
\usepackage{subfig}

\usepackage{amsfonts}
\usepackage{pifont}

\usepackage{color} 
\definecolor{purple}{RGB}{160,32,240}
\usepackage {comment}
\usepackage{url}

\newcommand{\rockstar}{\textsc{Rockstar}}
\newcommand{\ctrees}{\textsc{Consistent Trees}}
\newcommand{\vmax}{v_\mathrm{max}}

\def\aap{\mbox{A\&A}}
\def\apjl{\mbox{ApJ}}
\def\apjs{\mbox{ApJS}}

\graphicspath{{./final_plots/}}

\def\msun{\mbox{M$_{\odot}$}}

\def\mvir{\mbox{$M_{\rm vir}$}}

\def\mstar{\mbox{$M_{\rm C}$}}
\def\mpeak{\mbox{$M_{\rm peak}$}}

\def\vmax{\mbox{$V_{\rm max}$}}

\def\hmpc{h^{-1}{\rm Mpc}}

\defcitealias{RDA12}{RDA12}
\newcommand{\lambdap}{\lambda_{\rm B}}
\newcommand{\lambdaP}{\lambda_{\rm P}}

\newcommand{\cnfw}{C_{\mathrm{NFW}}}
\newcommand{\cklypin}{C_{\mathrm{Klypin}}}
\newcommand{\rvir}{R_{\mathrm{vir}}}
\newcommand{\rs}{R_{s}}
\newcommand{\pvir}{P_{\rvir}}
\newcommand{\amhalf}{a_{M_{1/2}}}
\newcommand{\almm}{a_{\mathrm{LMM}}}
\newcommand{\prho}{\mathcal{P}(\rho_{\sigma})}
\newcommand{\xoff}{X_{\mathrm{off}}}
\def\ltsima{$\; \buildrel < \over \sim \;$}    % Use in text mode
\def\lesssim{\lower.5ex\hbox{\ltsima}}           % Use in math mode
\def\gtsima{$\; \buildrel > \over \sim \;$}    % Use in text mode
\def\grtsim{\lower.5ex\hbox{\gtsima}}           % Use in math mode
\title[Properties of Dark Matter Halos: Local Environment Density]{Properties of Dark Matter Halos as a Function of Local Environment Density}

\author[]{Christoph T. Lee,$^{1}$\thanks{E-mail: chtlee@ucsc.edu}, Joel R. Primack$^1$, Peter Behroozi$^2$, Aldo Rodr\'iguez-Puebla$^3$, 
\newauthor Doug Hellinger$^1$, Avishai Dekel$^4$ \\
$^{1}$ Physics Department, University of California, Santa Cruz, CA 95064, USA\\
$^2$ Astronomy and Physics Departments and Theoretical Astrophysics Center, University of California, Berkeley, Berkeley CA, 94720, USA \\
$^3$ Instituto de Astronom\'ia, Universidad Nacional Aut\'onoma de M\'exico, A. P. 70-264, 04510, M\'exico, D.F., M\'exico \\
$^4$ Center for Astrophysics and Planetary Science, Racah Institute of Physics, The Hebrew University, Jerusalem 91904, Israel \\
}

\date{Accepted Dec 20 2016. Received Dec 15 2016; in original form Oct 6 2016}

\pubyear{2016}

\begin{document}

\label{firstpage}
\pagerange{\pageref{firstpage}--\pageref{lastpage}}

\maketitle

\begin{abstract}
We study how properties of discrete dark matter halos depend on halo environment, characterized by the mass density around the halos on scales from 0.5 to 16 $\hmpc$.  We find that low mass halos (those less massive than the characteristic mass $M_{\rm C}$ of halos collapsing at a given epoch) in high-density environments have lower accretion rates, lower spins, higher concentrations, and rounder shapes than halos in median density environments.  Halos in median and low-density environments have similar accretion rates and concentrations, but halos in low density environments have lower spins and are more elongated.  Halos of a given mass in high-density regions accrete material earlier than halos of the same mass in lower-density regions.  All but the most massive halos in high-density regions are losing mass (i.e., being stripped) at low redshifts, which causes artificially lowered NFW scale radii and increased concentrations.  Tidal effects are also responsible for the decreasing spins of low mass halos in high density regions at low redshifts $z < 1$, by preferentially removing higher angular momentum material from halos.  Halos in low-density regions have lower than average spins because they lack nearby halos whose tidal fields can spin them up.  We also show that the simulation density distribution is well fit by an Extreme Value Distribution, and that the density distribution becomes broader with cosmic time.
\end{abstract}

\begin{keywords}
Cosmology: Large Scale Structure - Dark Matter - Galaxies: Halos - Methods: Numerical
\end{keywords}

%%%%%%%%%%%%%%%%%%%%%%%%%%%%%%%%%%%%%%%%%%%%%%%%%%

%%%%%%%%%%%%%%%%% BODY OF PAPER %%%%%%%%%%%%%%%%%%

\section{Introduction}

In the $\Lambda$CDM standard modern theory of structure formation in the universe, galaxies populate dark matter halos and subhalos.  The properties of these halos and their distributions in space are therefore important in understanding the properties and distribution of galaxies.  The present paper investigates how the properties of the dark matter halos correlate with their environments, in particular with the mass density on various scales around the halos.  We also try to understand the origins of these correlations, studying in particular how halo environments affect halo evolution.

The effects on dark matter halos of their environments have been studied in many earlier papers.  Since the earliest cold dark matter paper \citet{BFPR}, it has been assumed that the initial conditions were Gaussian, as predicted by the simplest cosmic inflation models, which permitted treatment of halo properties based on analysis of the linearly evolved initial conditions.  The early N-body simulations \citep[e.g.,][]{DEFW95} had resolution too low to permit identifying dark matter halos so that galaxies had to be identified with individual particles in the simulations, which led to some misleading conclusions such as the supposed need for high bias.  But improved analysis of the initial conditions, for example by the ``peaks'' approach \citep{Kaiser84,PeacockHeavens85,BBKS}, permitted more detailed treatments of dark matter halo clustering \citep[e.g.,][]{Dalal+08,DesjacquesSheth10}.  Combination of the peaks approach plus N-body simulations led to further insights, including that halos in dense regions that do not correspond to high peaks in the initial conditions accrete more slowly than halos that do correspond to high peaks \citep{LudlowPorciani11}.  

Faster supercomputers and better codes have led to improved N-body simulations, in which the halos and subhalos that host galaxies, groups, and clusters are resolved.  Such simulations have permitted more detailed analyses of the correlations of halo properties with each other and with the halo environments measured in various ways.  One of the first papers to do this was \citet{LemsonKauffmann99}, which used $\Omega_0=1$  $\tau$CDM and $\Omega_0=0.3$ $\Lambda$CDM simulations and concluded that only the halo mass distribution varies as a function of environment, with more high mass halos in denser environments, in reasonable agreement with the analytic calculation by \citet{MoWhite96} based on extended Press-Schechter theory and the spherical top-hat model.  While subsequent N-body calculations confirmed this density effect on the halo mass function, they implied that environmental density also affects halo major merging rates \citep{Gottloeber+01} and other halo properties.  \citet{Bullock+2001} found that halos in dense environments tend to have higher concentrations than isolated halos.  \citet{ShethTormen04}, using the same simulations as \citet{LemsonKauffmann99}, found evidence that low-mass halos form somewhat earlier in dense environments.  \citet{Avila-Reese+99} found that dark matter halos that are isolated or in intermediate density environments have outer density profiles $r^{-\beta}$ with $\beta \approx 2.8 \pm 0.5$ while halos in denser regions have a wider range of $\beta$ up to $\sim 5$.  \citet{Avila-Reese+05}, again using $\Lambda$CDM simulations that were very small by modern standards, nevertheless found that halos in dense regions had lower spin parameters and higher concentrations and were less prolate than halos in lower density environments, with the differences arising mostly at low redshifts $z<1$ from phenomena such as tidal stripping in dense environments.  In the present paper, we confirm and expand on these results.  \citet{Maulbetsch+07} used a $512^3$ particle $\Lambda$CDM simulation in a $(50 \hmpc)^3$ volume to study the mass accretion history of galaxy-mass halos in different environments, and found that halos of the same final mass accreted their mass earlier in denser environments, and also accreted a significantly higher fraction of their mass in major mergers.  They suggested that this could help to explain the galaxy density-morphology relation, that early type galaxies are more common in dense environments.  They also found that $\sim80\%$ of halos in higher density environments are not accreting (i.e., have $\dot{M} \leq 0$), while this fraction is only $\sim20\%$ in low density environments.  They defined the environmental density on a scale of $4 \hmpc$, with density less than average considered ``low'' and density greater than 6 times background density considered ``high,'' but they found similar results measuring density on scales from 2 to $8\hmpc$.  They also found that subtracting the central halo mass in determining the environmental density, as \citet{LemsonKauffmann99} had done, made little difference to the results.  In the present paper we do not subtract the central halo mass in determining the environmental density, since we consider densities in volumes much larger than those of their central halos {\color{black} \citep[i.e. for a given halo mass, we only consider environment densities smoothed on scales greater than $4 \ \rvir$; see \S5.1 for further explanation; cf.][]{Muldrew+12}}.

Measuring environmental density in spheres of radii 1, 2, 4, and $8 \hmpc$ in a set of relatively small $\Lambda$CDM simulations in various volumes, \citet{Maccio+07} found that higher-concentration low-mass halos are found in denser environments, and lower-concentration ones in less dense environments. This is consistent with the higher concentration of early-forming halos \citep{Bullock+2001,Wechsler+02}; the earlier formation of low-mass halos in dense regions \citep[e.g.,][]{ShethTormen04}; and the finding that high-concentration low-mass halos are more correlated than average \citep[i.e., more biased:~][]{GaoSpringelWhite05,Wechsler+2006,GaoWhite07}, a phenomenon that has become known as ``assembly bias.''  
Here,  low-mass halos means those less massive than the characteristic mass $M_{\rm C}$ of halos collapsing at the present epoch \citep[see, e.g.,][Fig. 9]{Paper1}.  

\citet{FaltenbacherWhite10}, analysing the Millennium simulation \citep{Millennium}, found that near-spherical and high-spin halos show enhanced clustering.
\citet{FakhouriMa10}, also analysing the Millennium simulation, showed that mergers are increasingly important for halo mass growth in denser regions while diffuse accretion dominates growth in voids \citep[elaborating on the results of][]{Maulbetsch+07}, with galaxy-mass halos forming somewhat earlier in denser environments, where they accrete less at low redshifts because the dark matter there has higher velocity dispersion \citep[as also argued by][]{WangMoJing07,Dalal+08}.  

\citet{WangMoJingYangWang11} studied properties of dark matter halos as a function of their environments, characterized mainly by the tidal field but also by density on a scale of $6 \hmpc$.  They found that high-density environments provide more material for halos to accrete, but the stronger tidal fields there tend to suppress accretion.  They found that halos in higher tidal field environments and in higher density environments have higher spins, with the trends stronger for higher mass halos.  

\citet{SkibbaMaccio11} and \citet{Jeeson-Daniel+11} used principal component analyses to study the correlations of many halo properties including environment.  \citet{SkibbaMaccio11} used the overdensity in spheres of 2, 4 and $8\hmpc$ to measure the environmental density, and found that at fixed halo mass the environmental density does not significantly determine any of a halo's properties.  \citet{Jeeson-Daniel+11} did a more detailed correlation study of halo properties.  The strongest correlation they found was between halo concentration and age, with more concentrated halos also being more spherical and having lower spin.  But they found that there was little correlation of halo properties with their environment measure.  Instead of using overdensity, they measured the environment using a quantity they call $D_{1,0.1}$, equal to the distance to the nearest friends-of-friends halo with a mass greater than 10\% of the halo's mass divided by the radius of the neighbour's halo.  Unlike overdensity, $D_{1,0.1}$ does not correlate with halo mass \citep{Haas+12}.  
 
%We do not leave out unrelaxed halos
Some authors \citep[e.g.,][]{Bett+07,Maccio+07,SkibbaMaccio11,Ludlow+12,Ludlow+13} have studied mainly dark matter halos that are ``relaxed'' according to various criteria, such as an upper limit on $D_{\rm off} = |{\bf r}_{\rm Peak}  -  {\bf r}_{\rm CM}|/r_{\rm vir}$, the offset of the halo density peak from the center of mass within the halo radius, in units of the halo radius, or on the virial ratio $T/|U|$.  In this paper we study all halos, not just ``relaxed'' ones, because all the halos of mass $\grtsim 10^{10} \msun$ will host at least one central galaxy, regardless of whether it is relaxed or not, and our main motivation for studying halo properties as a function of environment is to clarify the implications for galaxies in environments of various densities. In this paper we also restrict attention to discrete halos (i.e., those that are not subhalos) since we are interested in using the dark matter halos to understand the properties of their central galaxies.

%Outline of this paper.
In the present paper we use the new $\Lambda$CDM Bolshoi-Planck simulation \citep{Klypin+16,Paper1} to study the environmental dependence of halo properties and their evolution.  The simulation is summarized in \S2, and we explain there how we measure the environmental density around every halo.  \S3 describes how the environmental density depends on the scale on which it is measured, and provides fitting functions using Extreme Value Distributions.  \S4 describes the environmental dependence of dark matter halo mass functions.  In \S5.1 we describe correlations between halo properties and environment at the present epoch. In \S5.2 we study the redshift evolution of halo properties at different densities, showing the origin at higher redshifts of the trends we found at $z=0$.  In the remaining subsections of \S5 we discuss the mass accretion rate (\S5.3),  halo concentration (\S5.4), halo spin (\S5.5), and halo prolateness (\S5.6).  \S6 summarizes and discusses our results.  The Appendix contains figures that expand upon issues discussed in the text or present alternative plots.

%Role of halo stripping summarized here, treated in detail in a companion paper.
We have found that a large fraction of lower mass halos in dense environments are stripped, that is they have less mass today than their main progenitors did at some earlier epoch.  We have found that halo stripping is the main cause of the decrease in spin and the increase in concentration of lower-mass halos in dense regions.
We discuss such effects of halo stripping briefly in the present paper, with more detailed results and discussion of the causes of halo stripping in a companion paper (Lee et al. 2016, in preparation).

%%%%%%%%%%%%%%%%%%%%%%%%%%%%%%%%%%%%%%%%%%%%%%%%%%%
\section{Simulations and Method}
\label{sec:Sims}

In this paper we use the  \rockstar\  halo finder \citep{ROCKSTAR} and \ctrees\ \citep{CTrees} to analyse results for the recent Bolshoi-Planck $\Lambda$CDM simulation, with $(2048)^3$ particles in a volume of $(250 \ \hmpc)^3$, based on the 2013 Planck \citep{PlanckC13} cosmological parameters $\Omega_M = 0.307$, $\Omega_B = 0.048$, $\sigma_8=0.823$, and $n_s=0.96$ and compatible with the Planck 2015 parameters \citep{PlanckC15}.  With a mass per particle of $1.5\times10^8 \ h^{-1} M_\odot$ and a force resolution of 1 kpc, the Bolshoi-Planck simulation has adequate resolution to identify halos that can host most visible galaxies.  This simulation, along with the larger MultiDark-Planck simulations, have been analysed in detail with fitting functions provided for many halo properties \citep[see especially][]{Paper1} and all of the halo catalogues and merger trees are publicly available.\footnote{\rockstar\ catalogues (including local densities around halos) and \ctrees\ merger trees used here are available at \url{http://hipacc.ucsc.edu/Bolshoi/MergerTrees.html}, and FOF and BDM catalogues are available at \url{https://www.cosmosim.org/cms/simulations/multidark-project/}.}   

% Gaussian smoothing on various length scales
	
\begin{figure*}
	\centering
	\includegraphics[trim=100 0 130 0, clip, width=\textwidth]{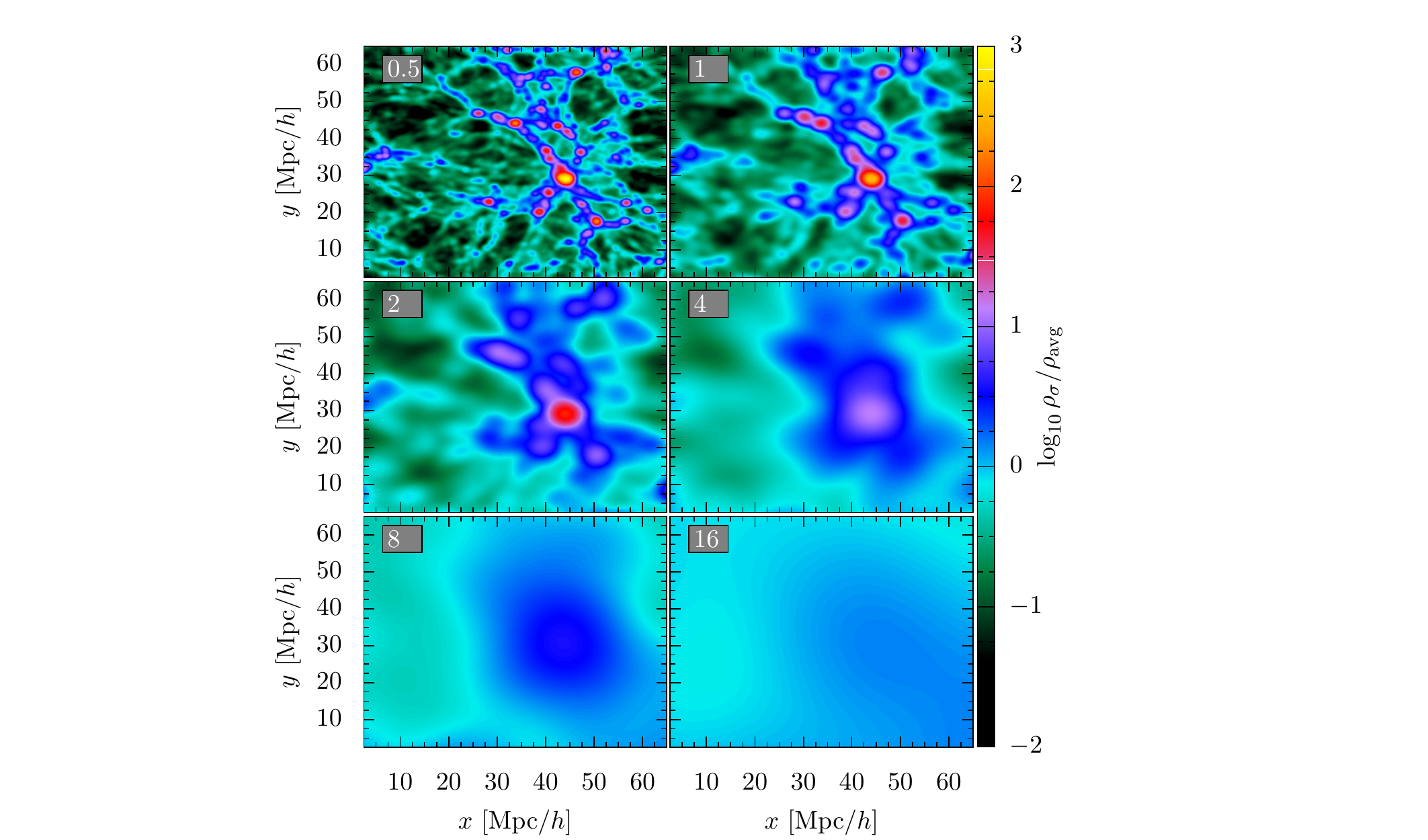}
    \caption{An example region of the Bolshoi Planck simulation at z = 0, coloured by local environment density smoothed on scales of $\sigma = 0.5$, 1, 2, 4, 8, and 16 $\hmpc$. Densities are reported with respect to the average density of the full volume.  The dynamic range of smoothed densities decreases at larger scales, dampening the color contrast between low and high density regions. Average density regions are highlighted by the light-blue to light-green color transition.  This example region has a depth of $1/4 \ \hmpc$ and a side length of $62.5 \ \hmpc$.}
    \label{fig:bp_z0_density_slice}
\end{figure*}

	We implement a Gaussian smoothing procedure to compute the density of the full simulation volume smoothed on many different length scales.  We convolve the CIC density cube with a 1-dimensional Gaussian kernel applied sequentially along each axis ($x$, $y$, $z$).  We refer to the half-width at half-maximum (HWHM) of the Gaussian kernel as the smoothing radius ($\sigma_{s}$), or smoothing scale.  The kernel ($k$) is truncated at cell $t$, where $k[t] < k[0]/10^{5}$, and then renormalized.  Since we use a voxel size of $250 \ [\hmpc] / 1024 \approx 1/4 \ \hmpc$, we've chosen to smooth the box on scales of $1/2$, $1$, $2$, $4$, $8$, and $16 \ \hmpc$.  We then update each halo in the \rockstar\ halo catalogues with the CIC and smoothed density values corresponding to their locations in the simulation volume.  {\color{black} We prefer using this method to other methods of determining local density (e.g. growing spheres of different sizes around halo positions) primarily because it is highly efficient and provides the density field at each point in the simulation volume.}  In Fig. \ref{fig:bp_z0_density_slice} we show the density smoothed on different scales for an example region of the simulation, with a depth of $1/4 \ \hmpc$ (1 voxel) and a height and width of $62.5 \ \hmpc$.  Average density regions are distinguished by light blue-green colouring, indicating the transition between underdense (green to black) and overdense (dark blue to yellow) regions.

%%%%%%%%%%%%%%%%%%%%%%%%%%%%%%%%%%%%%%%%%%%%%%%%%%%%%%
\section{Density Distributions}
\label{sec:Density_Distributions}

 % Explain how the density distribution plots were made and what they represent
 
\begin{figure*}
    \centering
    \subfloat[][]{\label{fig:fitted_liny} \includegraphics[trim=5 15 25 25, clip, width=\columnwidth]{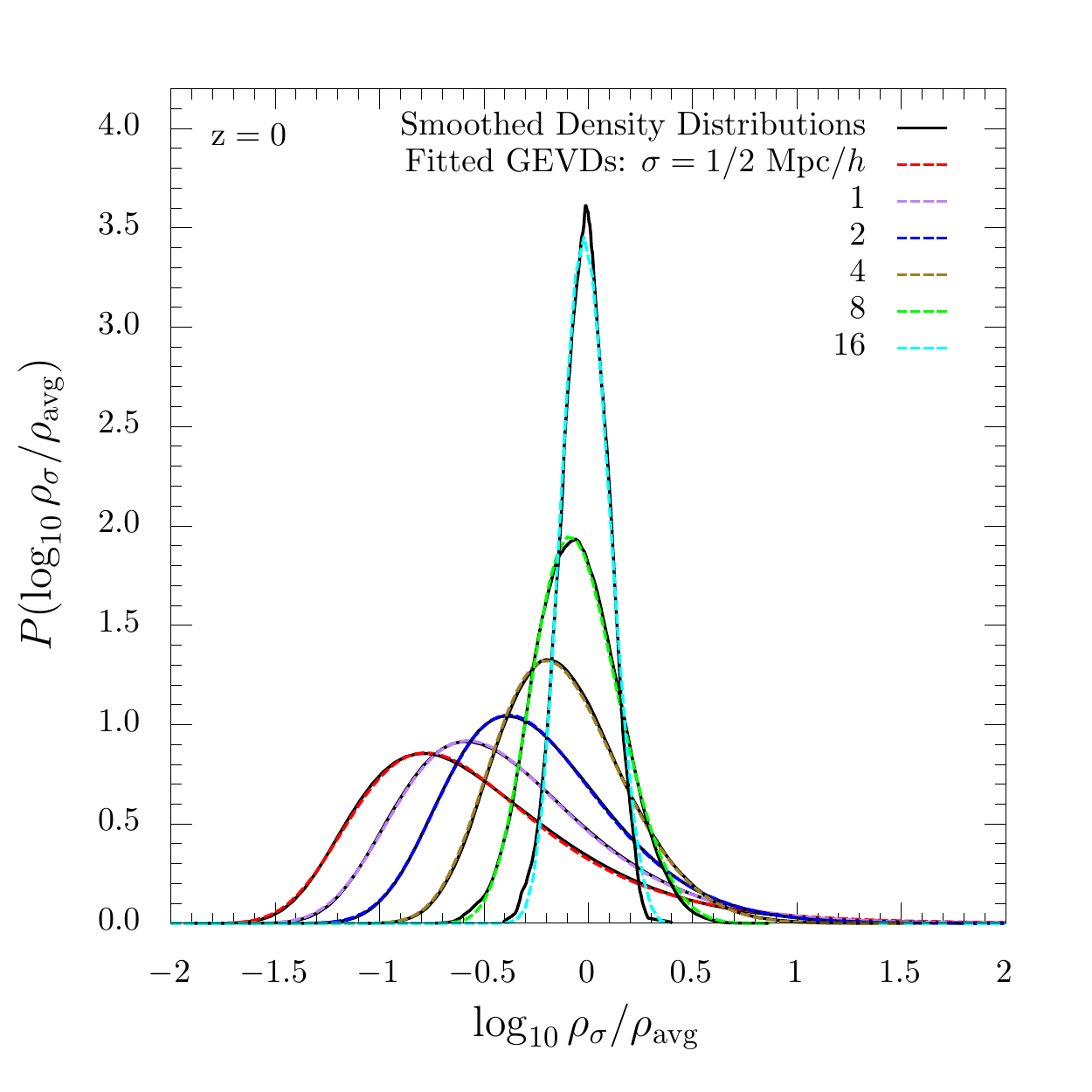}}
    \qquad    \subfloat[][]{\label{fig:fitted_logy} \includegraphics[trim=5 15 25 30, clip, width=\columnwidth]{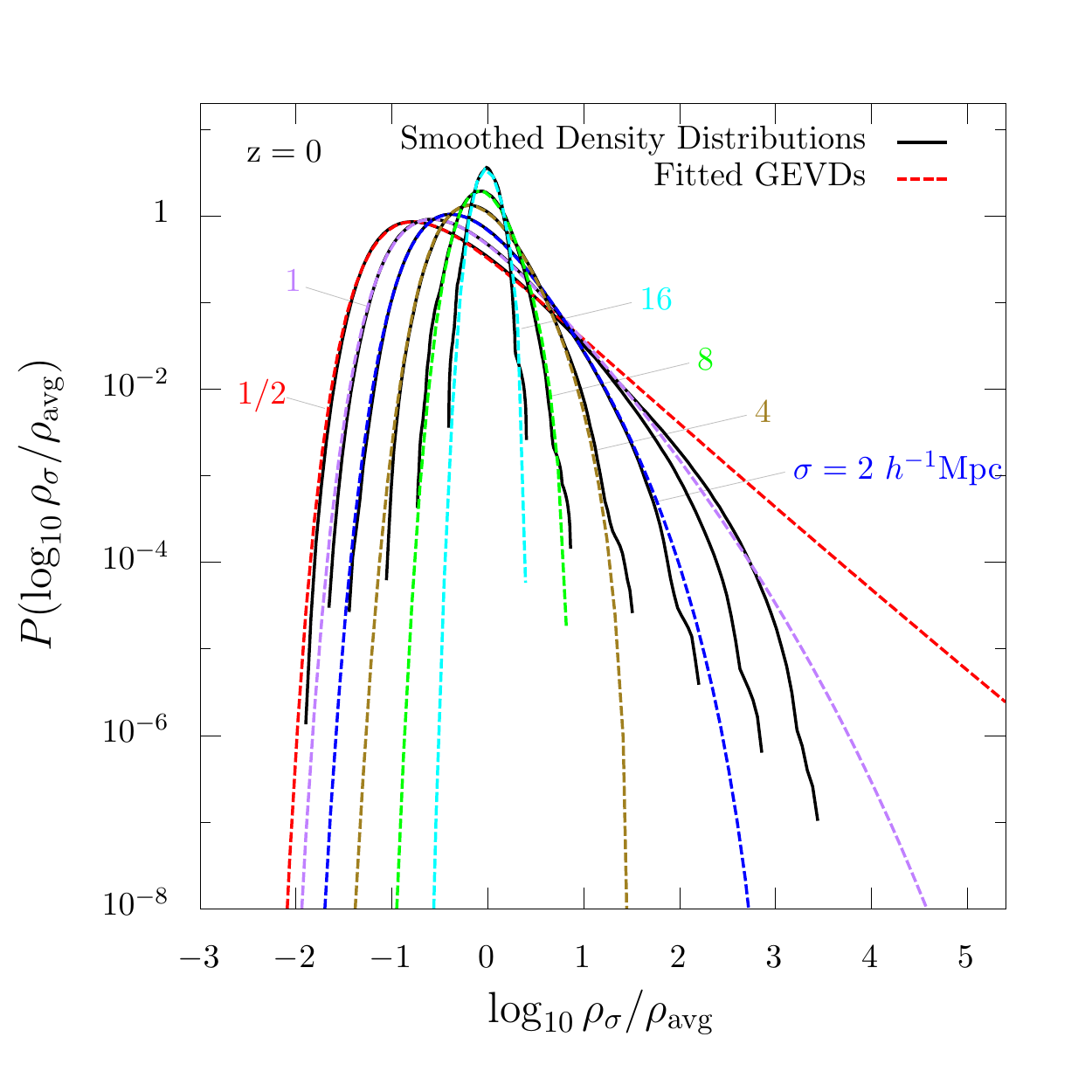}}
    \caption{Probability distributions of local environment density for the entire simulation volume at $z = 0$, shown with linear \protect\subref{fig:fitted_liny} and log \protect\subref{fig:fitted_logy} scaling on the vertical axis.  Data is shown in black solid lines, while analytical fits are shown with coloured dashed lines. Labels refer to the smoothing length (HWHM) of the Gaussian kernel used on the CIC density voxelization.  Densities are reported with respect to the average density.  The density distributions are well fit by the Generalized Extreme Value distribution, with small smoothing scales resembling a Gumbell type distribution, and larger smoothing scales resembling a Weibull type distribution.  Non-linear structures (in particular, voids) are relatively less abundant on larger scales.}
    \label{fig:density_dist}
\end{figure*}

	In Fig. \ref{fig:density_dist}, we present probability distributions of densities smoothed on several length scales for the Boshoi-Planck simulation at redshift $z=0$.  We compute these distributions using the full simulation volume, with black lines indicating different smoothing scales, and coloured lines representing best-fitting analytic functions for each smoothing scale.  We report densities with respect to the average density of the simulation to clearly distinguish between underdense and overdense regions.  The smallest scales probe the widest range of densities, from the centres of voids to the centres of massive halos.	Extreme values in the density field are redistributed over a larger volume when smoothed on larger scales, reflected in the narrower total range of densities observed for larger smoothing scales.  The shapes of the distributions indicate the abundance of non-linear structures present at a given length scale.  Densities smoothed on scales $\sigma \geq 8 \ \hmpc$ have a nearly log-normal distribution peaked around average density.   This indicates that density fluctuations on these scales are dominated by large waves in the Gaussian random field, with weak contributions from non-linear structures (voids, etc.) with radii greater than $8 \ \hmpc$.  On smaller scales ($\sigma \leq 2 \ \hmpc$),  underdense regions contain more volume than overdense regions, indicating that these scales are probing at or below the scales of non-linear structures like cosmic voids and filaments.  At very high densities, the distributions become noisy due to poor statistics and sensitive to the voxelization of halo cores, especially for smaller smoothing scales.

\begin{figure}
	\includegraphics[trim=5 15 25 28, clip, width=\columnwidth]{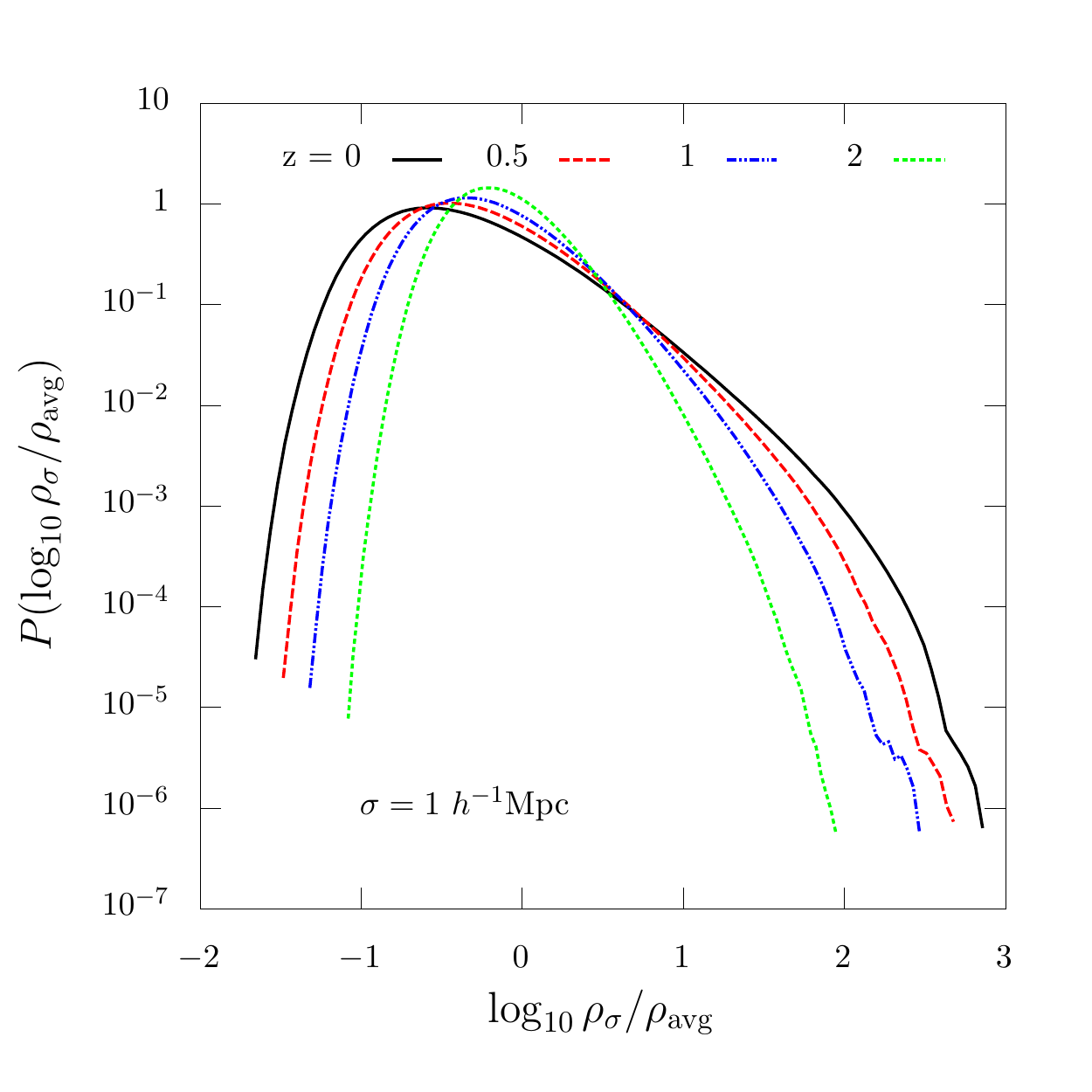}
    \caption{Probability distributions of local environment density smoothed using $\sigma_{s} = 1 \ \hmpc$ for the entire simulation volume, shown with log scaling on the vertical axis. Different coloured lines represent the same smoothing scale, but at different redshifts.  Non-linear structure emerges more dramatically at lower redshifts.  Voids grow emptier, while filaments and clusters grow denser with time.}
    \label{fig:density_distr_log_z}
\end{figure}

\begin{figure*}
	\centering
	\includegraphics[trim=80 120 145 130, clip, width=0.9\textwidth]{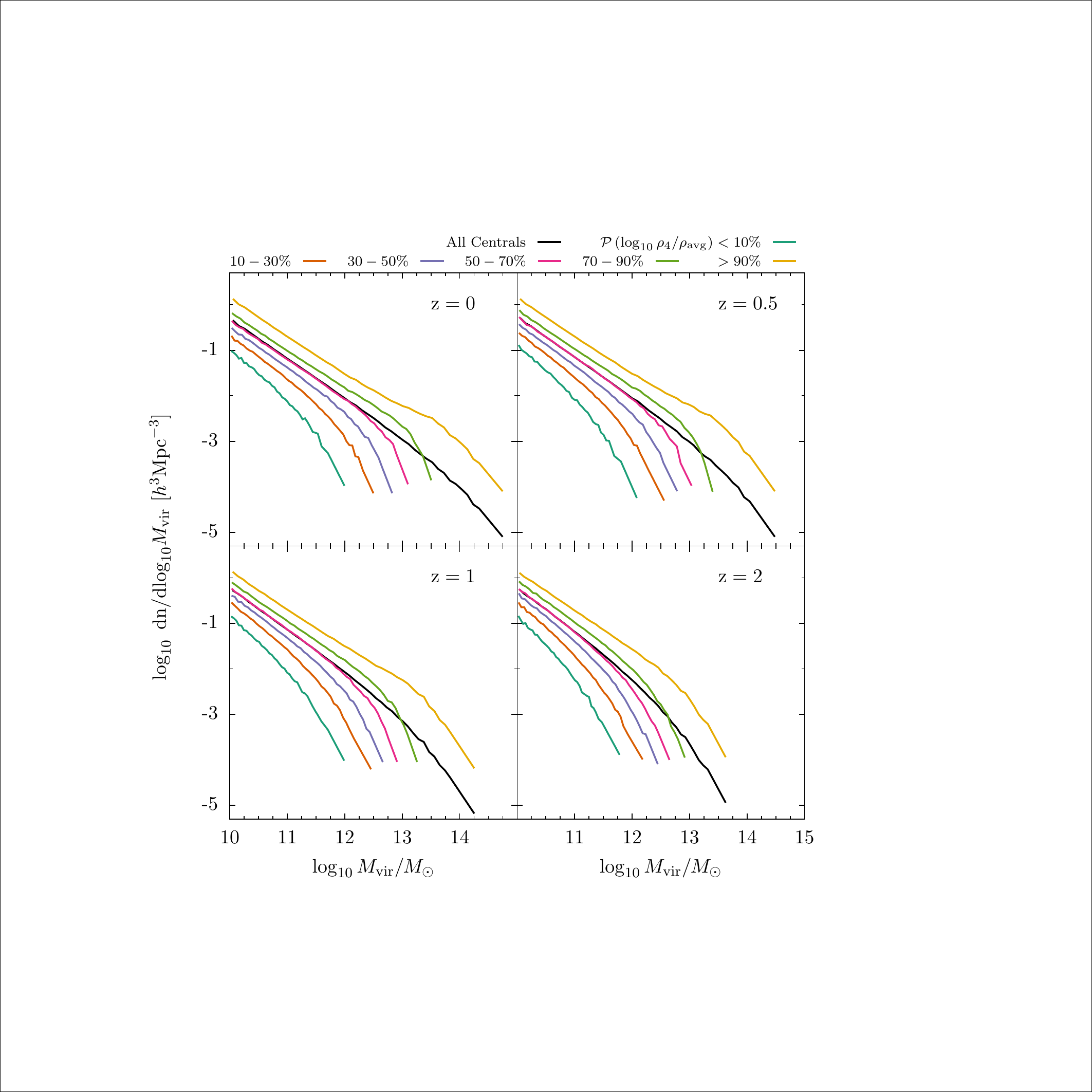}
    \caption{Halo mass functions at redshifts 0, 0.5, 1, and 2.  Coloured lines represent halo mass functions computed in percentile bins of local environment density smoothed with $\sigma = 4 \ \hmpc$, while black lines indicate the mass function of all distinct halos.  In order to consistently distinguish high from low density regions at different redshifts, we determine density percentiles relative to the whole simulation volume, rather than just the locations of the halos (e.g. $\mathcal{P} < 10\%$ reflects the lowest density voxels in the entire simulation, most of which are probably voxels in voids that contain no halos).  We see that each mass function has a characteristic mass, above which the abundance drops off more rapidly.  This characteristic mass is lowest in low density regions and highest in high density regions.  We find that at $z = 2$, the characteristic masses are lower and cover a narrower range of masses compared to at $z = 0$.  Additionally, at $z = 2$ the slopes of the mass functions change more gradually from below to above the characteristic masses compared to at $z = 0$.  The mass function of halos in the highest density regions is also somewhat steeper above the characteristic mass compared to at $z = 0$.  These differences reflect the flow of dark matter in the simulation: that voids become emptier and clusters become richer with time.}
    \label{fig:z0_density_hmf}
\end{figure*}

\subsection{Generalized Extreme Value Distribution}
\label{sec:GEVD}

 % Explain what the GEVD is, what fitting technique we used, what the implications of the fit are
 
 	We find that the probability distributions of smoothed cosmic densities are well fit by a Generalized Extreme Value (GEV) distribution, as defined in Eq. \ref{eq:genevd}, where $x$ is a random variable, $\beta$ is a scale parameter, $k$ is a shape parameter, and $\mu$ is a location parameter.  Extreme value theory \citep[see, e.g.,][]{ExtremeValueDistributions,ExtremeValueTheory}
describes the statistics of extrema of samples drawn from random distributions, and has been applied in analysis of 2D and 3D cosmological datasets \citep{Colombi+11}.
\begin{eqnarray}
    f(x) = \frac{1}{\beta}\mathrm{exp}\left[-\left(1+kz\right)^{-1/k}\right]\left(1+kz\right)^{-1-1/k}
    \label{eq:genevd} , \\
    z = \frac{x-\mu.}{\beta}
	\label{eq:genevd_z}
\end{eqnarray}
The GEV distribution encompasses several sub-families known as the Gumbell, Weibull, and Fr\'echet distributions, distinguished by shape parameter $k = 0$, $k < 0$, and $k > 0$, respectively.  In Table \ref{tab:fitted_params} we provide best fit values of $\beta$, $k$, $\mu$, and the residual sum of squares (RSS) for each density distribution.  The RSS is defined as  
\begin{equation}
RSS = \sum_{n=0}^{m} \left[f\left(x[n],\beta,k,\mu\right) - P\left(x[n]\right)\right]^{2},
\label{eq:rss}
\end{equation}
where $f(\cdots)$ is a GEV distribution with parameters $\beta$, $k$, and $\mu$, $x[n]$ are the binned density values and $P(x[n])$ is the real density distribution, as shown by the black lines in {\color{black} Fig. \ref{fig:density_dist}}. We use a simulated annealing fitting algorithm to minimize the RSS between the fit and the data.

We find that for small smoothing scales ($\sigma = 0.5$ and $1 \ \hmpc$), the distribution functions are well approximated by a Gumbell type distribution ($k \approx 0$), while larger smoothing scales ($\sigma = 2$, $4$, $8$, and $16 \ \hmpc$) are better approximated by a Weibull type distribution ($k < 0$).  {\color{black}Furthermore, we note that $k \propto \log \sigma$ (i.e. $k$ decreases nearly constantly in log space from $\sigma = 0.5$ to $16 \ \hmpc$, with the largest decrease occurring between $\sigma = 0.5$ and $1 \ \hmpc$).} The location parameter $\mu$, corresponding to the peak in the distributions, changes from $\mu \sim -1$ for small smoothing scales to $\mu \sim 0$ for larger scales, reflecting the shift in the abundances of voids on those scales.  The scale parameter, correlating with the width of the distributions, decreases from $\beta \sim 0.5$ when smoothed on $1/2 \ \hmpc$ scales to $\beta \sim 0.1$ for $16 \ \hmpc$ scales.
\begin{table}
	\centering
	\caption{Best fit values to 3-parameter generalized extreme value distribution (Eqns. \ref{eq:genevd}, \ref{eq:genevd_z}).  $\sigma$ is the HWHM smoothing length, $\beta$ is the scale parameter, $k$ is the shape parameter, $\mu$ is the location parameter, and $RSS$ is the residual sum of squares between the fit and the data.}
	\label{tab:fitted_params}
	\begin{tabular}{lcccr} % four columns, alignment for each
		\hline
		$\sigma$ [$h^{-1}$Mpc] & $\beta$ & $k$ & $\mu$ & $RSS$\\
		\hline
		0.5 & 0.43 & 0.077 & -0.78 & 0.0035\\
		1 & 0.40 & -0.049 & -0.60 & 0.0012\\
		2 & 0.35 & -0.099 & -0.42 & 0.00083\\
		4 & 0.28 & -0.16 & -0.25 & 0.0026\\
		8 & 0.19 & -0.19 & -0.13 & 0.095\\
		16 & 0.11 & -0.24 & -0.051 & 1.47\\
		\hline
	\end{tabular}
\end{table}

\subsection{Evolution with Redshift}
\label{sec:Density_Distribution_Evolution}

 % Explain meaning / significance of the redshift evolution plot 

	We additionally calculate probability distributions of smoothed densities at redshifts $z = 0.5$, $1$, and $2$.  In Fig. \ref{fig:density_distr_log_z} we show the evolution of the distribution of densities smoothed with $\sigma = 1 \ \hmpc$ in comoving units (the simulation maintains a comoving volume of $[250 \ \hmpc]^{3}$ at each time step).  Fig. \ref{fig:density_distr_log_z_sigma4} shows the distribution of densities smoothed with $\sigma = 4 \ \hmpc$.
Non-linear structure emerges more dramatically at low redshifts, evidenced by the increasingly asymmetric peak at lower densities. Generally, we see that voids become less dense and higher density regions become more dense with time.  Material in voids empties into walls, filaments, and nodes, which grow ever denser. 

We note that \citet{Colombi+97} found a fitting function for the probability distribution of density using perturbation theory, and \citet{Sheth98} found that an Inverse Gaussian Distribution provided a good fit to the distribution of densities from simulations with white noise initial conditions.   \citet{Valageas+04} developed a model for the evolution of the density probability distribution function.  However, our GEV fits appear to be more accurate.

%%%%%%%%%%%%%%%%%%%%%%%%%%%%%%%%%%%%%%%%%%%%%%%%%%%%%%%
\section{Environmental Dependence of Halo Mass Functions}
\label{sec:HMF_density}

% Describe selection criteria for mass functions in different density regions and ramifications

	We present halo mass functions at redshifts $z = 0$, 0.5, 1 and 2 for halos in different density regions in Fig. \ref{fig:z0_density_hmf}.  We define consistent density ranges at each redshift by selecting percentile ranges of the smoothed CIC density values for the entire simulation volume{\color{black}, where coloured lines are computed in percentile bins of local environment density smoothed with $\sigma =  4 \ \hmpc$, while black lines indicate the mass function of all distinct halos.}  For example, $\mathcal{P} = 45-55\%$ selects halos with local densities equal to the median density of the entire simulation volume (including the densities of regions without halos).  These selection criteria have the advantage of providing an intuitive and consistent definition for different redshifts, while still sampling a fairly wide range of halo local densities.  Nevertheless, due to evolution in the distribution of local densities (Fig. \ref{fig:density_distr_log_z}), the population counts in these density percentile ranges change with redshift.  The distribution of halo local densities does not evolve in the same way that the whole volume does, since the halo distribution is mass-weighted rather than volume-weighted (i.e. we are not following evolution of voxels without halos).  In general, the distribution of halo densities is more similar to the full volume density distribution at higher redshifts and for larger smoothing scales, and less similar at lower redshifts and smaller smoothing scales.  As redshift decreases, halos tend to move towards higher density percentiles relative to the full volume densities. These trends are illustrated in Fig. \ref{fig:density_cdf_cdf}.  {\color{black}Still, we feel this approach is more consistent than using percentile bins determined from halo densities alone, since it involves no choice of halo mass range.  Choosing percentile ranges from halo densities would first require the selection of a halo mass range to use, which would have an arbitrary (user-defined) redshift dependence.}

% Describe trends observed in mass functions and implications

		We find that each density range has a characteristic mass, above which the abundance of halos falls more rapidly.  This characteristic mass increases monotonically with density, and is highest in the highest density regions.  The characteristic mass for a given density range increases with decreasing redshift, and is highest at $z = 0$.  However, the characteristic mass range (from the highest to lowest density regions) is narrower at high redshift than at low redshift.  The slope of the mass function above the characteristic mass for halos in the highest density regions at $z = 2$ is also steeper than at $z = 0$.  These differences reflect differences in the evolution of the mass-weighted halo density distribution relative to the volume-weighted full volume density distribution (Fig. \ref{fig:density_cdf_cdf}).  At $z = 0$, we are probing a more extremal population of halos in the very highest and very lowest density regions compared to at $z = 2$.  This effect is more pronounced in the highest density population than the lowest density population.

% Mention companion projects to compare simulated halo mass functions to observed galaxy luminosity functions

		Naturally, we are interested to know how well these halo mass functions in different density environments agree with observational data.  We have several projects underway to address this topic.  Using data from the Galaxy and Mass Assembly (GAMA) survey, \citet{McNaught-Roberts+14} determined galaxy local densities by counting surrounding galaxies within a sphere of $8 \ \hmpc$, and used them to compute galaxy luminosity functions in regions of different density. In ongoing work, our group is using abundance matching to compute luminosity functions in regions of different density, and comparing with data from the Sloan Digital Sky Survey (SDSS) and GAMA survey.  In a different approach, we are using void density profiles from SDSS cosmic void catalogues \citep[see, e.g.][]{Sutter+12} to assign local densities to galaxies in low-density regions of the SDSS, which we can then use to test predictions from our simulations.

%%%%%%%%%%%%%%%%%%%%%%%%%%%%%%%%%%%%%%%%%%%%%%%%%%%%%%%%
\section{Correlations with Local Environment Density}
\label{sec:density_correlations}

\subsection{Correlations at the present epoch}

\begin{figure*}
	\centering
	\includegraphics[trim=21 10 85 10, clip, width=\textwidth]{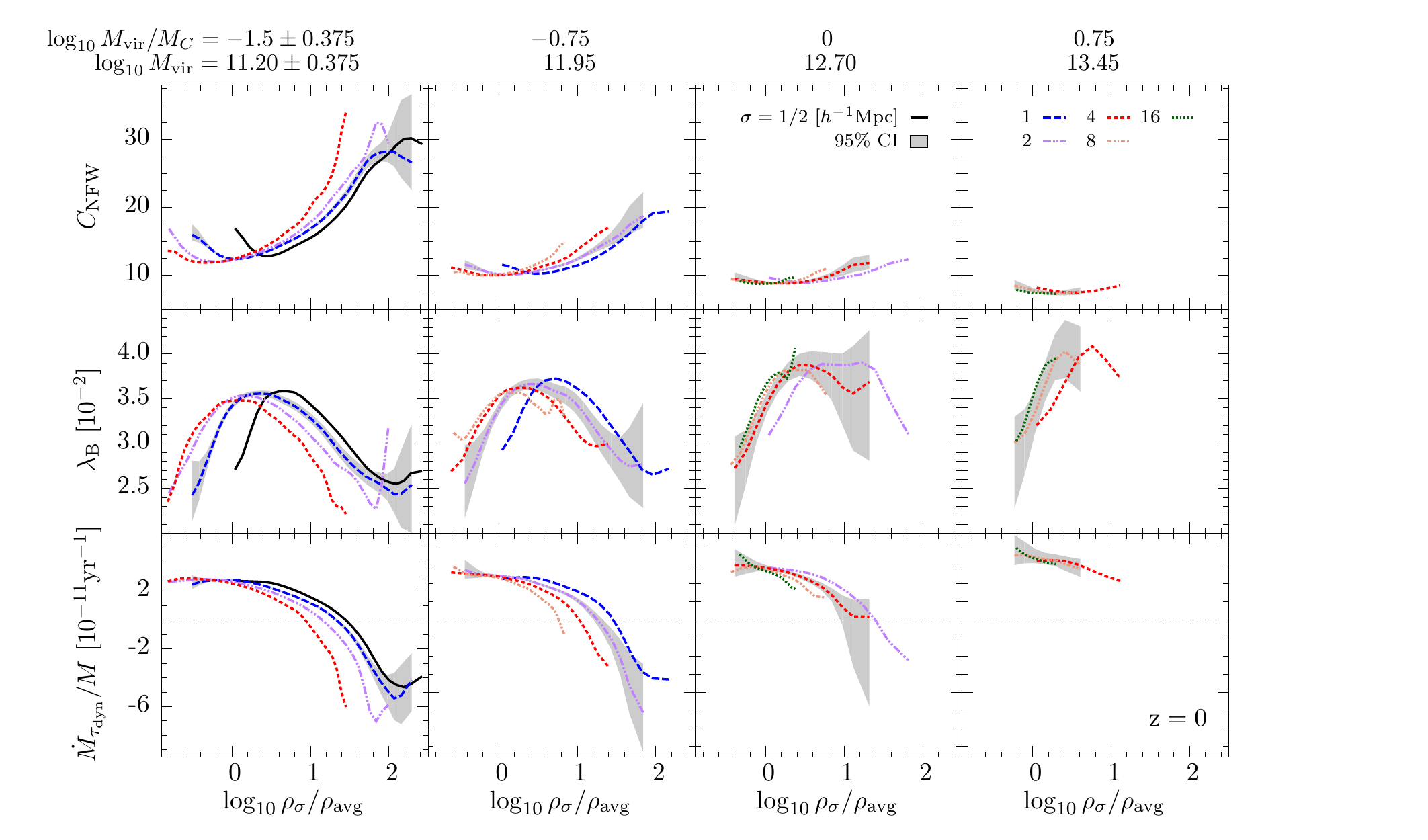}
    \caption{Medians of scatter in $\rho-\cnfw$, $\rho-\lambdap$, and $\rho-\dot{M}/M$ relationships at $z = 0$, where $\rho_{\sigma}$ is the local environment density smoothed on different scales and $\rho_{\mathrm{avg}}$ is the average density of the simulation.  Different coloured lines represent different smoothing scales.  The shaded grey filled curve represents the 95\% confidence interval on the median, shown only for the characteristic smoothing length $\sigma_{s,\mathrm{char}} = 1, 2, 4$, and $8 \hmpc$ for mass bins from left to right, respectively, and provides an indication of sample size at different densities.  Mass bins are selected relative to the non-linear mass ($\log_{10} \mstar = 10^{12.7} \msun$ at $z = 0$) to facilitate comparison between halos above, at, or below $\mstar$.  We see that lower mass halos occupy regions with a wide range of local densities, while higher mass halos are restricted to higher density regions.  Note also that larger smoothing scales will shift the range of densities towards the average density, so equal smoothing lengths should be used to compare density ranges for halos of different masses.  See Fig. \ref{fig:z0_correlations_p} for a discussion of the trends seen in this plot.}
    \label{fig:z0_correlations}
\end{figure*}

\begin{figure*}
	\centering
	\includegraphics[trim=21 10 85 10, clip, width=\textwidth]{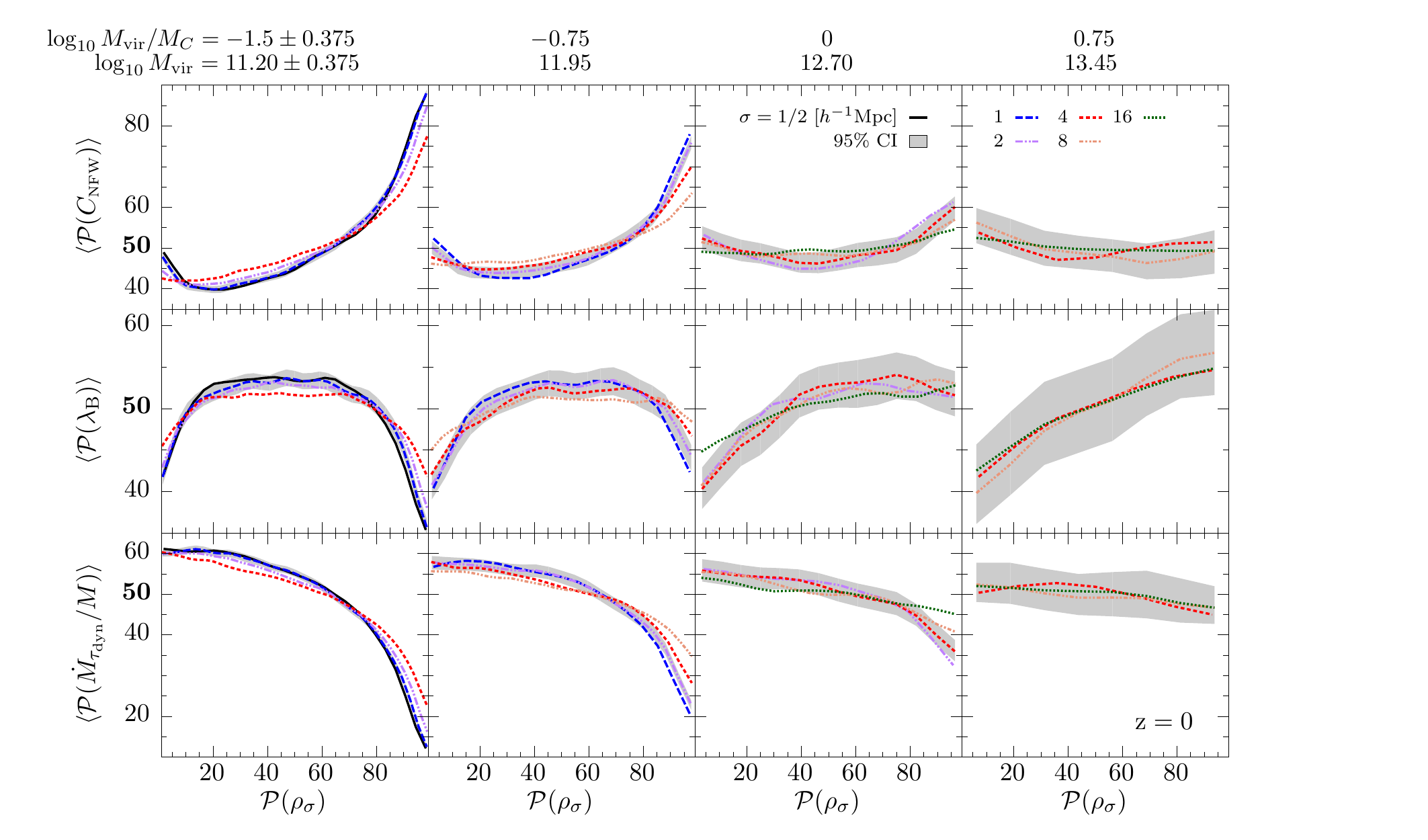}
    \caption{Medians of scatter in rank ordered distributions of $\rho-\cnfw$, $\rho-\lambdap$, and $\rho-\dot{M}/M$ at $z = 0$.  The vertical axes reflect the rank ordered percentile of the medians in each density percentile bin, with respect to all distinct (central) halos in the given mass bin.  The shaded grey filled curve represents the 95\% confidence interval on the median, shown only for the characteristic smoothing length $\sigma_{s,\mathrm{char}} = 1, 2, 4,$ and $8 \hmpc$ for mass bins from left to right, respectively.  For halos less massive than $\mstar$, we see that concentrations tend to be lower in lower density regions, except at the lowest densities, where they increase.  Spin parameters are maximized in median density regions, and decreased in high and low density regions.  Accretion rates are heavily suppressed in high density regions and maximized in low density regions.  For halos with $\mvir \geq \mstar$, the trends are less dynamic and less well constrained due to low statistics, but show similar relationships overall.  Note that high mass halos are not found in low density regions, so the trends observed represent trends in relatively high density regions only.}
    \label{fig:z0_correlations_p}
\end{figure*}

% Explain how density correlation plots were made (+percentilization)

	We turn our attention now to correlations between halo properties and local density at $z = 0$. In Fig. \ref{fig:z0_correlations}, we show medians of correlations between local density parameters and NFW concentration ($\cnfw$), Bullock spin parameter ($\lambdap$), and dynamically time-averaged specific mass accretion rate ($\dot{M}/M$).  Each column reflects a different halo mass range, while each row reflects a different halo property.  We chose mass bins such that we could compare halos below, at, and above the characteristic mass $\mstar$ of halos that are collapsing at $z = 0$, which is $10^{12.70} \msun = 5.0  \times 10^{12} \msun$ for the Planck cosmological parameters used in the Bolshoi-Planck simulation \citep[see][Figure 9, for a plot showing $\mstar$ as a function of redshift]{Paper1}.  For each panel, different lines represent different smoothing radii used to determine the local density around halos.  Due to the close relationship between virial radius and halo mass, we've plotted only smoothing radii that are greater than $\sigma_{s,min} \sim 4 \rvir$.  We then define the characteristic smoothing length as $\sigma_{s,char} = 2 \sigma_{s,min}$, which provides a reliable picture of local effects, while not being too much influenced by a halo's own profile or averaged out by larger smoothing scales.  These characteristic smoothing radii are, from left to right columns, $\sigma_{s,char} = 1$, $2$, $4$, and $8 \ \hmpc$, respectively.  {\color{black}To avoid overly crowding plots with lower mass halos, we plot only smoothing radii from $\sigma_{s} = \frac{1}{2} \sigma_{s,char}$ to $4 \sigma_{s,char}$.}  The curves with corresponding grey shading reflect the $95 \%$ confidence interval on the median for the characteristic local density for a given mass bin.
	
	In addition to choosing smoothing radii sufficiently greater than the virial radii of the halos, we balance each mass bin to have a flat mass-density relation.  For a given mass bin, this involves 2 dimensional sub-binning by halo mass and a given local density parameter, then randomly eliminating halos from appropriate sub-bins to force approximately equivalent mass distributions for each density sub-bin (limited by the coarseness of the 2D sub-grid).  We perform this procedure uniquely for each mass bin - density smoothing radius pair used.  If this procedure is not done, the results would be contaminated by an underlying mass dependence, which is noticeable at smaller smoothing radii but rather insignificant for larger radii.  Finally, we smooth the relations using a Gaussian filter with $\sigma = b(\mvir)$, where $b(\mvir)$ is the size of the horizontal (density) bins in a given mass bin.  So, regions with wider density bins are smoothed with a larger $\sigma$ than regions with narrow density bins.  We determine bin widths based on the number of halos in the population under consideration, with better statistics allowing for smaller bins.

	The resulting relations in Fig. \ref{fig:z0_correlations} show some clear density dependence at lower masses, where the range of local densities probed is high, while the data at high masses is comparatively flat and spans a narrower range of densities.  The significantly extended range of densities home to lower mass halos reflects that these halos may be found in regions both underdense (voids) and very overdense (clusters), while more massive halos tend to reside exclusively in higher density regions.  Additionally, lines representing different smoothing radii are shifted relative to each other, due to the averaging out of extreme densities with increasing smoothing scale.
	
	In order to more effectively analyse these correlations, we prepared an alternate representation using percentilized axes.  In Fig. \ref{fig:z0_correlations_p}, in a given mass bin, we've rank ordered halos by density parameter and the plotted halo property.  The resulting curves represent the percentile ranks of the medians (determined relative to the entire mass bin) of each halo property for a given density percentile rank.  This representation has the advantage of shifting and stretching the curves on each panel to facilitate comparison between different smoothing scales and halo masses.  We also provide Fig. \ref{fig:z0_density_cdf} as a means of translation between Figs. \ref{fig:z0_correlations} and \ref{fig:z0_correlations_p}, by relating actual values of halo properties to corresponding percentile ranks.  This percentilized form of correlations between halo properties and local density will be the basis for much of our ensuing discussion.

% Explain basic trends we see in percentilized density correlation plot

	In Fig. \ref{fig:z0_correlations_p}, we see that except in the lowest density regions, low mass halos ($\mvir < \mstar$) have median concentrations that scale monotonically with increasing local density.  Surprisingly, we also find that low mass halos in the lowest $15\%$ of local densities have higher concentrations than halos in the roughly $20-40$th percentile range.  So, for halo masses less than the characteristic mass $M_{\rm C}$, we find halo concentration scales strongly with local density, with the caveat that concentrations go up in very low density regions.  Halos at or above $\mstar$ display a much weaker correlation between density and concentration, though massive halos tend to be more concentrated in lower density regions.  For $\lambdap$, we find that halos less massive than $\mstar$ in both high and low density regions have lower spin parameter compared to halos in median density regions.  More massive halos, however, tend to have spin parameters that scale monotonically with local density.  Lastly, all halos tend to accrete less in higher density environments, though low mass halos exhibit far stronger accretion suppression than massive halos.  Interestingly, this indicates that halos in low density regions (bottom $20\%$ of densities) accrete more rapidly than halos in higher density regions.
		
% Explain how progenitor history plots were made
\subsection{Redshift evolution of halo properties at different densities}

\begin{figure*}
	\centering
	\includegraphics[trim=18 154 60 160, clip, width=\textwidth]{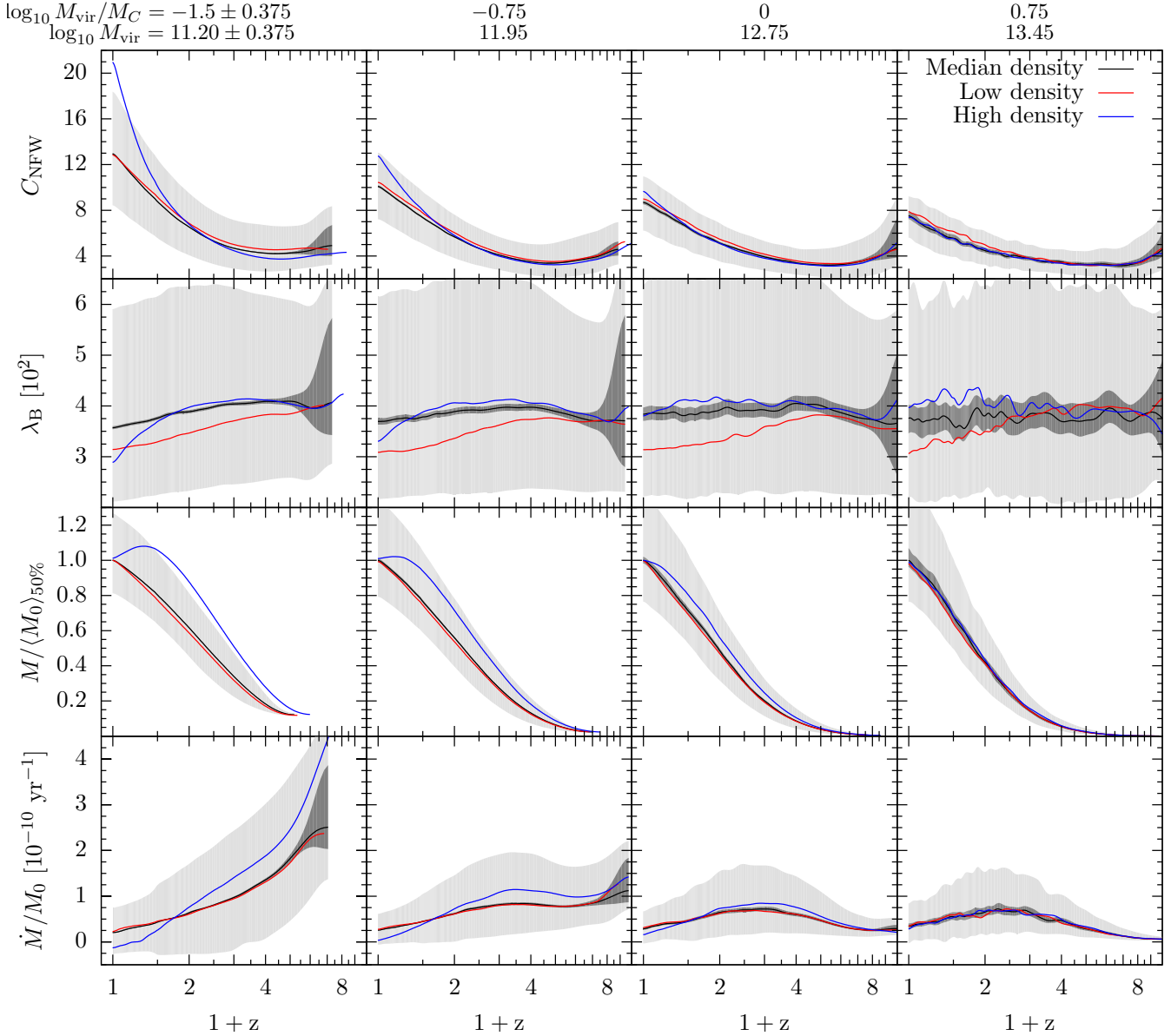}
    \caption{NFW Concentration, spin parameter, mass, and specific mass accretion rate histories for halos that end up in high (blue), median (black), and low (red) density regions at z = 0.  Halos are selected based on their percentile rank in characteristic local density parameter ($\sigma_{s,\mathrm{char}} = 1, 2, 4,$ and $8 \hmpc$ for mass bins from left to right, respectively).  Halos that are in the percentile ranges $\mathcal{P} = 0-10, 45-55,$ and $90-100$ represent halos in low, median, and high density regions, respectively.  The curves reflect median properties of the progenitors of the z = 0 halo populations.  The dark grey shading reflects the 95\% confidence interval on the median and the light grey shading reflects the $20-80\%$ dispersion of each property, shown only for halos in median density regions.  We see that low mass halos in high density regions at $z = 0$ experienced rapid growth of concentration and reduction of spin parameter at late times compared to halos in lower density regions.  Halos in high density regions also experienced sharp accretion rate suppression and even mass loss at late times.  Halos in low density regions at $z = 0$ had slightly higher concentrations and consistently lower spin parameters than halos in median density regions throughout most of their history.  Halos in low density regions accreted slightly less at early times and slightly more at late times compared to halos in median density regions.} 
    \label{fig:z0_density_prog_hist_1}
\end{figure*}

	One of the principal analysis methods we've used to investigate the origins of the trends in Fig. \ref{fig:z0_correlations_p} is to examine the median evolution of halo properties along the most massive progenitor branch (MMPB) of halos in regions of different density at $z = 0$.  In Figs. \ref{fig:z0_density_prog_hist_1} and \ref{fig:z0_density_prog_hist_2}, for a given mass bin, we've selected all halos in the $0-10$th, $45-55$th, and $90-100$th percentile ranges of characteristic local density $\sigma_{s,char}$ at $z=0$ to represent halos in low, median, and high density regions, respectively.  Using the halo merger trees, we follow the MMPB of each halo and record the properties of each progenitor.  We then present the median halo properties of the most massive progenitors of halos that end up in these low, median, and high density regions at $z = 0$.  Note that because the density selections are made at $z = 0$, the progenitors of those halos are not guaranteed to reside in similar density regions at higher redshifts.  Once an MMPB mass drops below the completeness threshold $M_{\mathrm{min}} = 10^{10} \msun$, we discard any remaining progenitors from the analysis.  This is done in order to exclude halos with low particle counts that may have unreliable halo properties.  The dark grey and light grey shading reflect the $95\%$ confidence interval on the median and the $20-80$th percentile range of the halo property at a given redshift, respectively.  These are shown only for halos in median density regions at $z = 0$, though similar trends apply to halos in low and high density regions at $z = 0$.
	
	In order to minimize bias introduced by the longest lasting MMPBs (those that remain above $M_{\mathrm{min}}$ out to higher than average redshifts), we implement a "median preserving" approach.  Tracing time backwards from $z = 0$, when a given MMPB drops below $M_{\mathrm{min}}$, we determine the halo property rank of that MMPB's earliest eligible progenitor $P_{\mathrm{earliest}}$ (with $\mvir > M_{\mathrm{min}}$) with respect to all other eligible halos at that time step.  Then, in addition to eliminating further progenitors of $P_{\mathrm{earliest}}$, we eliminate progenitors of the MMPB with rank $R^{\prime} = N-R$, where N is the total number of halo progenitors in consideration at the relevant redshift and R is the rank of $P_{\mathrm{earliest}}$.  For example, if the earliest eligible progenitor $P_{i}$ of a given MMPB ranks in the $67$th percentile in $\cnfw$ compared to all other eligible progenitors at that redshift, then in addition to eliminating the remaining progenitors of $P_{i}$, we eliminate any remaining progenitors of the MMPB that ranks in the $100-67 = 33$rd percentile in $\cnfw$ at that same redshift.  This procedure is applied uniquely for each halo property presented and provides a less biased determination of the median properties of halo progenitors than a simple low statistics cut-off.  {\color{black}For example, if we naively plot the median mass evolution of a group of halos, excluding progenitors that fall below $10^{10} \msun/h$, the median mass of the group will not fall below that threshold until every single halo in the group has done so.  Using the median preserving approach, however, the median halo mass would fall below the threshold sooner, because high mass progenitors would be paired off and removed along with low mass progenitors.}  Finally, each curve is smoothed using a Gaussian filter with smoothing $\sigma = 2$, $3$, $6$, and $10$ time steps for mass bins from left to right columns, respectively.  We discuss our interpretation of Fig. \ref{fig:z0_density_prog_hist_1} and related progenitor property figures in the following sections.

\subsection{Mass Accretion Rate}
\label{sec:mar}

% Recap trends observed in main density-correlation plots

	{\color{black}In Figs. \ref{fig:z0_correlations} and \ref{fig:z0_correlations_p}, we use the dynamically time averaged specific mass accretion rate, defined as
\begin{equation}
\frac{\dot{M}_{\tau_\mathrm{dyn}}}{M} = \frac{1}{M(t)}\frac{M(t) - M(t-\tau_{\mathrm{dyn}})}{\tau_{\mathrm{dyn}}},
\label{eq:mdotdyn}
\end{equation}
while in Fig. \ref{fig:z0_density_prog_hist_1} we use the instantaneous specific mass accretion rate, defined at a given timestep $t_{i}$, as 
\begin{equation}
\frac{\dot{M}}{M} = \frac{1}{M(t_{i})}\frac{M(t_{i}) - M(t_{i-1})}{t_{i} - t_{i-1}}.
\label{eq:mdotinst}
\end{equation}
See \citet{Paper1} Appendix A for details about the timesteps saved in the Bolshoi-Planck simulation we are using.} In Fig. \ref{fig:z0_correlations_p} we see that except for the most massive halos, those in higher density regions have suppressed accretion rates compared to halos in lower density regions. From Fig. \ref{fig:z0_correlations}, we can additionally see that in very high density regions, the median accretion rates for halos less massive than $\mstar$ even become substantially negative, indicating a net loss, or "stripping" of material from the halo.  Accretion rates of halos in median and low density regions tend to be very similar, if not marginally higher in low density regions.  It should again be noted that halos more massive that $\mstar$ span a much narrower range of local densities and have far poorer statistics compared to less massive halos, resulting in typically less dynamic correlations between halo properties and local density.
	
% Discuss points related to accretion rates of halos in high density regions at z = 0	
	
	Fig. \ref{fig:z0_density_prog_hist_1} Rows 3 and 4 show the evolution of virial mass and instantaneous specific accretion rate for halos in low, median, and high density regions at $z = 0$.  Halos in high density regions tend to have much more massive progenitors compared to halos in median and low density regions and accrete material more rapidly and sooner than halos in lower density regions. Halos in high density regions also experience much lower accretion rates at late times.  

\begin{figure*}
	\centering
	\includegraphics[trim=18 154 60 160, clip, width=\textwidth]{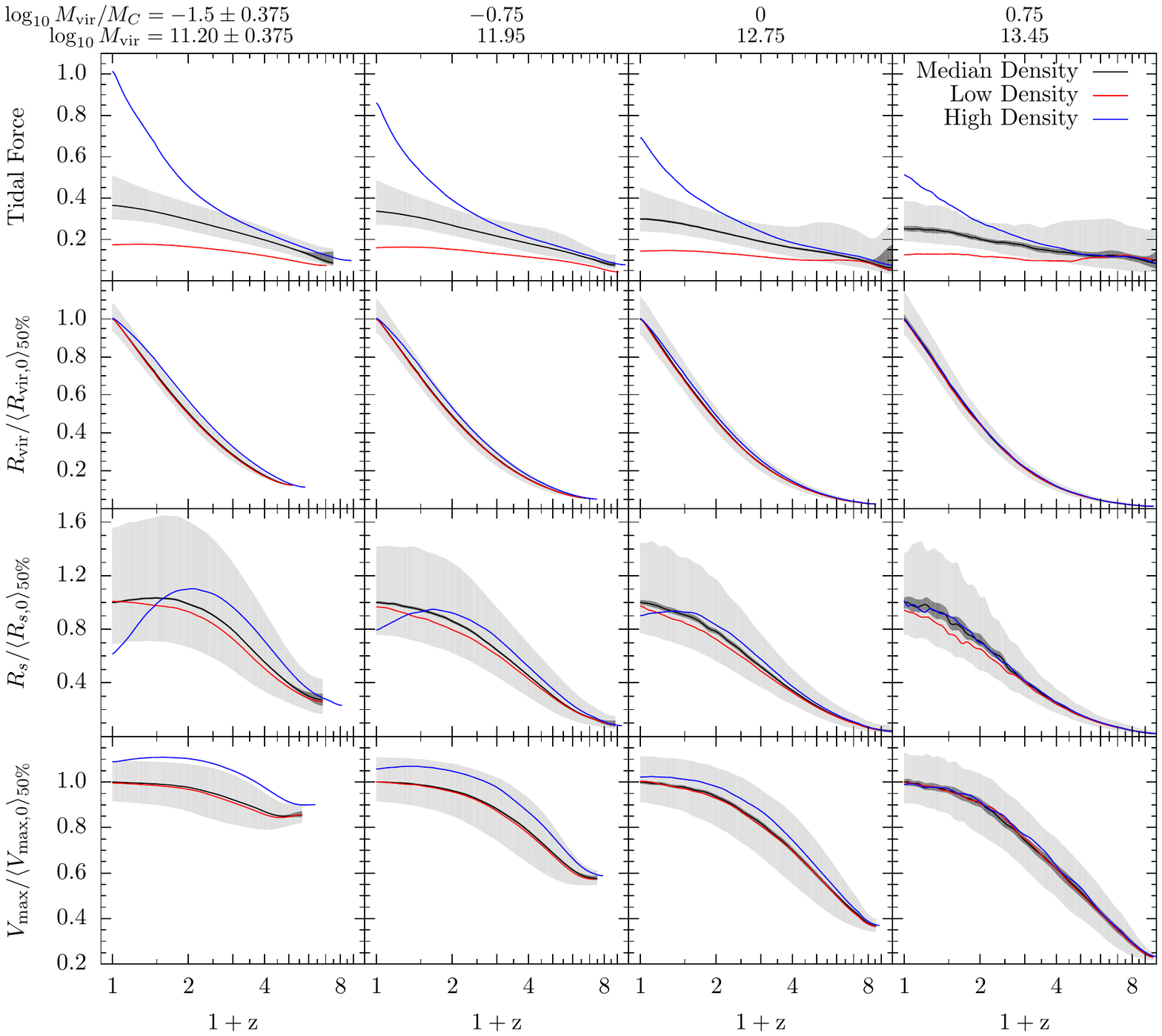}
    \caption{Same as Fig. \ref{fig:z0_density_prog_hist_1}, but showing tidal force, virial radius ($\rvir$), scale radius ($\rs$), and maximum circular velocity ($\vmax$).  In order to efficiently compare different mass bins, we normalize $\rvir, \rs,$ and $\vmax$ by the median values of the median density population at $z = 0$.  We see that halos in high density regions at $z = 0$ experience strong tidal forces at late times, but significantly weaker tidal forces at higher redshifts.  Since tidal force correlates strongly with local density, it seems halos in high density regions at $z = 0$ migrated from roughly median density regions around $z \lesssim 2$.  Halos in high density regions at late times evolved from halos with larger $\rvir$ and $\rs$ and higher $\vmax$ compared to halos in lower density regions, but experienced a dramatic reduction in scale radius at late times.  Halos in low density regions at $z = 0$ experienced consistently low tidal forces throughout their evolution and somewhat lower scale radii than halos in median density regions.}
    \label{fig:z0_density_prog_hist_2}
\end{figure*}

\begin{figure*}
	\centering
	\includegraphics[trim=21 10 85 10, clip, width=\textwidth]{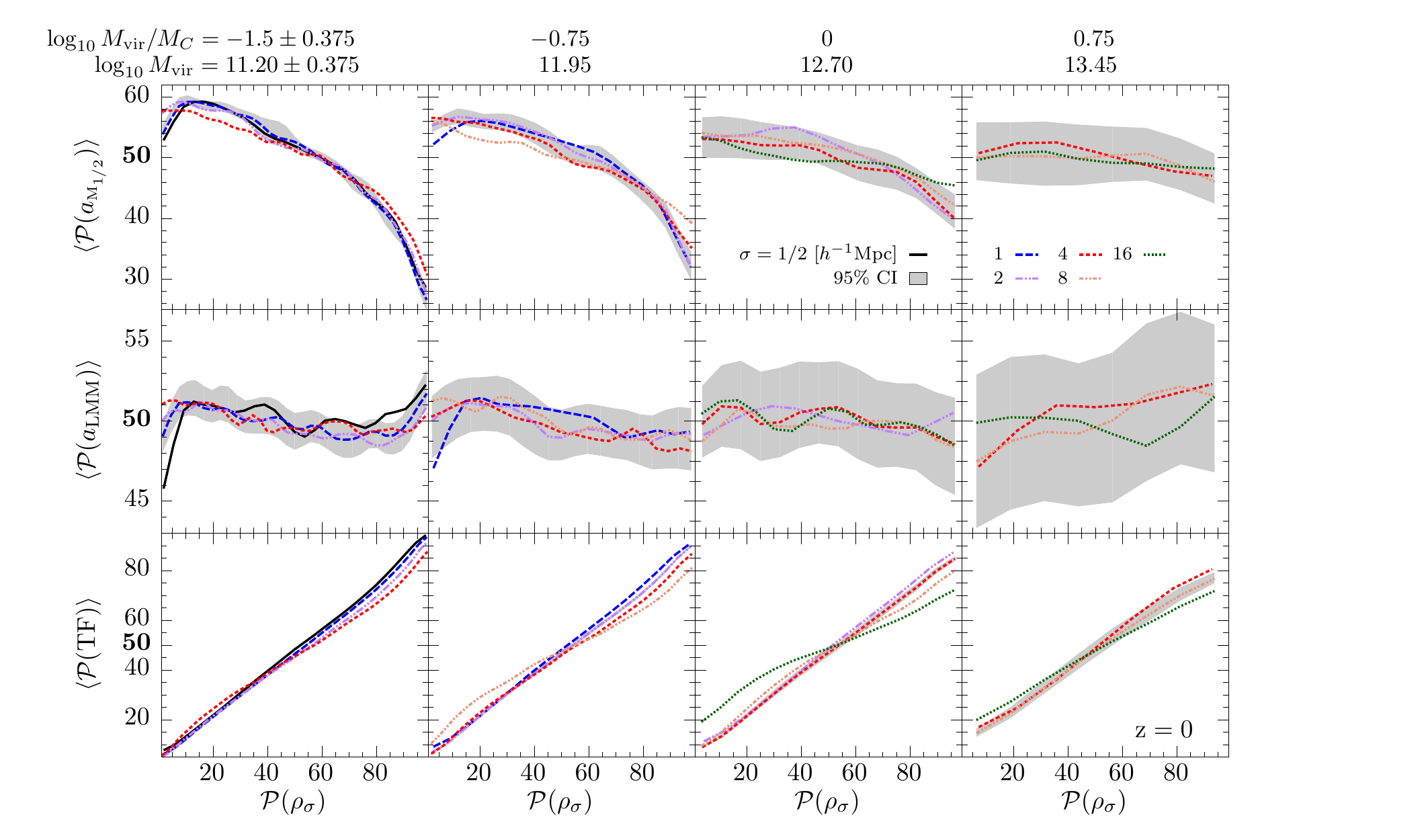}
    \caption{Same as Fig. \ref{fig:z0_correlations_p}, but showing half-mass scale factor ($\amhalf$), scale of last major merger ($\almm$) and tidal force.  We see that the percentilized $\amhalf-\rho$ relation is roughly inversely proportional to the $\cnfw-\rho$ relation.  Low mass halos in high density regions typically formed earlier than halos in lower density regions, except in the lowest density regions, where we see a downturn in $\amhalf$.  Low mass halos in very low density regions most recently experienced major mergers at earlier times than halos in higher density regions.  We observe little correlation between $\almm$ and local density above $\prho \approx 20$.  Tidal forces correlates strongly with local density, suggesting we can use tidal force history as a reliable tracer of local density history.}
    \label{fig:z0_correlations_p_2}
\end{figure*}

In Fig. \ref{fig:z0_correlations_p_2} Row 1, we plot the relation between local density and the half-mass scale factor $\amhalf$, used in this analysis as an indicator of halo formation time.  Consistent with the mass growth profiles from Fig. \ref{fig:z0_density_prog_hist_1}, we see that low mass halos in high density regions form earlier than halos in lower density regions.  In Fig. \ref{fig:z0_density_prog_hist_2} Row 1, we see that halos in high density regions at $z = 0$ experience much higher tidal forces than halos in lower density regions, particularly at $z \lesssim 1$.  Furthermore, {\color{black}Fig. \ref{fig:z0_correlations_p_2} Row 3} shows that tidal force correlates strongly with local density.  Thus, tidal forces histories along the MMPBs of $z = 0$ halos are probably closely related to their local density histories. For halos in high density regions at $z = 0$, the strongly reduced median tidal force their progenitors experience at redshifts $z \grtsim 1$ indicates that those progenitors live in closer to median density regions at earlier times, and only enter very high density regions at late times ($z < 1$).  Halos in high density regions at $z = 0$ thus form earlier and accrete significant amounts of material sooner than halos in lower density regions, however migration into very high density regions at $z < 1$ results in strongly reduced accretion rates and even net mass loss among lower mass halos.  Halos in high density regions, corresponding to regions with strong tidal fields, have suppressed accretion rates due to the reduced impact parameters infalling material would require to be captured by the halo (Hahn et al. 2009, Hearin et al. 2015).  Additionally, halos may lose weakly bound particles or those whose orbits are sufficiently elongated.
	
% Discuss points related to accretion rates of halos in low density regions at z = 0

	Halos in lower density regions tend to have higher accretion rates than halos in higher density regions, though this trend plateaus for halos in regions of less than median density.  One may expect accretion rates to be reduced in low density regions, due to the presumably decreased amount of local material available for accreting, but we find this is not the case.  In fact, in Fig. \ref{fig:z0_density_prog_hist_1} Row 3, we see that halos in low density regions evolve from slightly less massive progenitors than halos in median density regions, implying that halos in low density regions must accrete more rapidly at late times to end up with the same $z = 0$ masses as halos in median density regions.  Indeed, in Fig. \ref{fig:z0_density_prog_hist_1} Row 4, we see that the accretion rate histories of these two halo populations are extremely similar, albeit with halos in low density regions accreting marginally less at high redshifts ($z \grtsim 1$) and slightly more at low redshifts than halos in median density regions (consistent with the mass growth profiles).   Halos in low density regions experience consistently lower tidal forces throughout their lifetimes than halos in median density regions.  This suggests that halos in low density regions may be able to accrete a larger fraction of available surrounding material than halos in median density regions due to decreased competition from neighbouring halos. 

\subsection{Concentration}
\label{sec:cnfw}

% Recap main trends seen in concentration-density correlation plot and questions raised

	{\color{black}In this paper, we use the \citet[NFW]{NFW96} radial density profile to define concentration,
\begin{equation}
\cnfw \equiv \rvir/\rs,
\label{eq:cnfw}
\end{equation}
where the scale radius $\rs$ is determined by fitting a halo to the NFW profile,
\begin{equation}
\rho_{\mathrm{NFW}}(r) = \frac{4\rho_{s}}{(r/\rs)(1+r/\rs)^{2}}.
\label{eq:nfw}
\end{equation}
For halos with less than 100 particles, we revert to using the Klypin concentration $\cklypin$ \citep[see, e.g.][]{Klypin+2011,Paper1}, which can be solved for numerically using a relationship between $\vmax$ and $\mvir$.  Most of our analysis focuses on halos with more than 100 particles, except for when we follow the progenitors of halos to high redshift (e.g. Fig. \ref{fig:z0_density_prog_hist_1}). However, for comparison with Fig. \ref{fig:z0_correlations_p}, we also provide relations between $\cklypin$ and local density in Fig. \ref{fig:z0_correlations_p_3}. We do not include halos below $\mvir = 10^{10} \ \hmpc$ ($\sim 70$ particles) in any part of our analysis.}

	From Fig. \ref{fig:z0_correlations_p} Row 1, we see that NFW Concentration ($\cnfw$) of low mass halos ($\mvir < \mstar$) scales monotonically with local density, except for in low density regions, where we see an upturn in concentration.  Higher mass halos ($\mvir \geq \mstar$) display little correlation between concentration and local density, though it should be noted that higher mass halos tend to be confined to higher density regions.  The fundamental questions raised here are: (1) why do low mass halos in high density regions have much higher concentrations; and (2) why do low mass halos in very low density regions ($\prho < 20$) have higher concentrations than those in somewhat higher density regions ($\prho \approx 20 - 50$)? In order to understand these trends, we now examine correlations between several additional halo properties and local density, as well as the time evolution of halo concentration and related halo properties in regions of different density.

% Discuss significance of additional density correlations (half mass scale factor, scale of last MM)

	First, we look at the correlation between halo formation time and local density.  We use the scale factor at which a halo first reached half its peak mass ($\amhalf$) as an indicator of formation time.  In Fig. \ref{fig:z0_correlations_p_2} Row 1, we plot the relation between half-mass scale factor and local density, using the same method and plotting styles as in Fig. \ref{fig:z0_correlations_p}.  Especially for lower mass halos, we see that the relation between half-mass scale factor and density is inversely proportional to the $\cnfw$-$\rho_{\sigma}$ relation, even for the lowest density regions.  This indicates that halos in higher density regions generally formed earlier than those in lower density regions, with the exception that halos in very low density regions ($\prho < 20$) typically formed somewhat earlier than those in slightly higher density regions ($\prho \approx 20$).  If we assumed a simple model for concentration evolution, such as $\cnfw \propto (1+z)^{-1}$ at fixed mass as in \citet{Bullock+2001}, then the inverse proportionality between the $\cnfw-\rho_{\sigma}$ and $\amhalf-\rho_{\sigma}$ relations would seem consistent; halos in high density regions have higher concentrations \textit{because} they formed earlier, and thus have had more time to grow $\cnfw$.  However, this conclusion assumes that halo concentration evolves at the same rate in regions of different density, which, as we will show, is not the case.
	
	We plot in Fig. \ref{fig:z0_correlations_p_2} Row 2 the relation between the scale factor of the last major merger ($\almm$) and local density.  Overall the correlation is very weak, except for low mass halos in very low density regions, where we see that halos typically last experienced a major merger earlier than halos in higher density regions.  The downturn at low densities in the $\almm-\rho_{\sigma}$ relation roughly corresponds to the downturn at low densities in the $\amhalf-\rho_{\sigma}$ relation, upturn in $\cnfw$, and downturn in $\lambdap$, indicating these may all be due to the same mechanism.

% Discuss trends in concentration progenitor history plot and questions raised / answered

	We now turn to the evolution of halo concentration in regions of different density.  In Fig. \ref{fig:z0_density_prog_hist_1} Row 1, we select groups of halos in high, median, and low density regions at $z = 0$ and plot the median concentrations of their progenitors at each time step.  What we immediately see is that low mass halos in high density regions have only recently ($z \lesssim 1$) diverged sharply in concentration compared to halos in median and low density regions.  In fact, we see that at higher redshifts ($z \grtsim 1$), the $z = 0$ high density halos had equal, if not slightly lower concentrations than halos that end up in median and low density regions at $z = 0$.  Clearly, low-mass halos in high density regions have drastically different rates of concentration growth than halos in lower density regions at late times.  
Halos in low density regions historically have slightly higher concentrations than those in median density regions, though overall the difference between the two is slight.  However, we should keep in mind that from Fig. \ref{fig:z0_correlations_p} Row 1 we see that the upturn in concentration for halos in very low density regions occurs around $\prho \approx 20$, so by comparing low density ($\prho = 5$) to median density ($\prho = 50$), we are somewhat obscuring the low density upturn. Nevertheless, there does seem to be a systematic difference between concentrations histories of halos in median and low density regions at z = 0.  These results redirect our questions as follows: (1) why do halos in high density regions at $z = 0$ exhibit drastically increased concentrations at late times compared to halos in lower density regions; and (2) why do halos in low density regions at $z = 0$ have systematically higher concentrations throughout most of their lifetimes?  

% Discuss additional progenitor history plots and their relevance, further questions raised (Rs, Rvir, M, Mdot/M, TF, Vmax)

	Our next step is to break down the concentration evolution by examining the evolution of virial radius $\rvir$ and scale radius $\rs$, which are related by $\cnfw = \rvir/\rs$.  In Fig. \ref{fig:z0_density_prog_hist_2} Rows 2 and 3, we plot the evolution of virial radius and scale radius for halos in regions of different density at $z = 0$.  We see that all halos have relatively similar virial radii (as one would expect, given that we have normalized the mass-density correlation within each mass bin).  Halos in high density regions had larger virial radii at earlier times, reflecting that they evolved from higher mass halos than the median and low density groups (see also Fig. \ref{fig:z0_density_prog_hist_1} Rows 3 and 4 for accretion history comparison).  However, the variation in virial radius between halos in different density environments does not significantly contribute to the large concentration discrepancy at late times.  This leaves the scale radius, which indeed shows drastically different evolution between halos in different density regions.  Low mass halos ($\mvir < \mstar$) in high density regions at $z = 0$ experience rapid scale radius growth at early times, followed by a plateau around $z = 1$, and a sharp decline at late times.  We see a similar growth and plateau trend in median and low density regions, but no significant decrease in scale radius.  Additionally, low mass halos in low density regions have consistently lower scale radii than halos in median density regions.  This systematic difference in scale radius growth rate between halos in median and low density regions at $z = 0$ suggests there may be differences in the accretion histories of the two populations, but as shown in Fig. \ref{fig:z0_density_prog_hist_1} Row 4, the instantaneous accretion rate histories are practically identical.  However, we note that the redshift at which the accretion rate of halos in high density regions crosses below the accretion rate of halos in median and low density regions roughly corresponds to the redshift at which the scale radii of halos in high density regions begins to decrease.
	
	Since the scale radius represents the location at which the spherically averaged density profile transitions from an inner $\rho \propto r^{-1}$ to an outer $\rho \propto r^{-3}$ dependence, the differences in scale radius evolution indicate fundamental differences in the evolution of halo density profiles in regions of different density.  So, do halos in high density regions really have scale radii that are shrinking ($r^{-1}$ core physically decreasing in size)?  One way to probe structural changes to the central regions of these halos is by looking at maximum circular velocity evolution $\vmax$, shown in Fig. \ref{fig:z0_density_prog_hist_2} Row 4.  The systematically higher $\vmax$ for halos in high density regions is a reflection of their more massive progenitors.  We see that low mass halos in high density regions at $z = 0$ do display slightly decreased $\vmax$ at late times, indicating some net loss of high energy particles from the central regions of these halos.  However, this is not a strong effect, and certainly not indicative of a drastically reduced scale radius.  This suggests that these halos likely have density profiles that are evolving away from the NFW profile, resulting in a poor NFW fit with artificially reduced $\rs$ values.  Indeed, most low mass halos in high density regions have outer density profiles that fall off faster than $r^{-3}$ \citep{Avila-Reese+99}, presumably due to tidal stripping of material from the outer regions of these halos.  Forcing an NFW fit to these halos will tend to produce artificially small $\rs$ values,  due compensation for the steep fall off of their outer density profiles.
	
% Discuss additional density correlation plots of relevance (Cklypin, C with no stripped, BSR)
% Note:  leaving out Cklypin discussion at this time

	As shown by the halo mass evolution plot (Fig. \ref{fig:z0_density_prog_hist_1} Row 3), low mass halos in high density regions have dramatically reduced $\mvir$ at late times, due to negative accretion rates (i.e., stripping of material from the halos).  Halos in high density regions tend to be much more stripped than halos in low density regions.  Furthermore, in Fig. \ref{fig:z0_correlations_p_ns}, we see that when halos that have lost more than 2\% of their peak mass are removed from the population, the median halo concentration sharply decreases in high density regions, but changes little in median and low density regions.  This suggests that inflated concentrations in high density regions are simply a consequence of the modified density profiles (and subsequently poorer NFW fits) of halos that are undergoing extreme mass loss.  While exceptionally high concentrations correlate strongly with halos in very high density regions, they are a poor and indirect descriptor of the properties of these halos.  It would be better to characterize these halos using a fitting function that properly fits their outer regions. We are currently investigating how to best characterize halo profiles in regions of different density and for halos that have been appreciably stripped.
	
	There are only small differences in concentration evolution between halos in low density regions and those in median density regions, suggesting that differences in the median formation times may be the main reason for differences in the median concentrations of these populations.  Low mass halos in low density regions typically formed earlier, and have had more time to increase their concentrations compared to halos in median density regions.  Early forming halos will also tend to have higher central densities and smaller scale radii than late forming halos.

% Recap on which questions answered, which remain **

\begin{figure*}
	\centering
	\includegraphics[trim=21 10 85 10, clip, width=\textwidth]{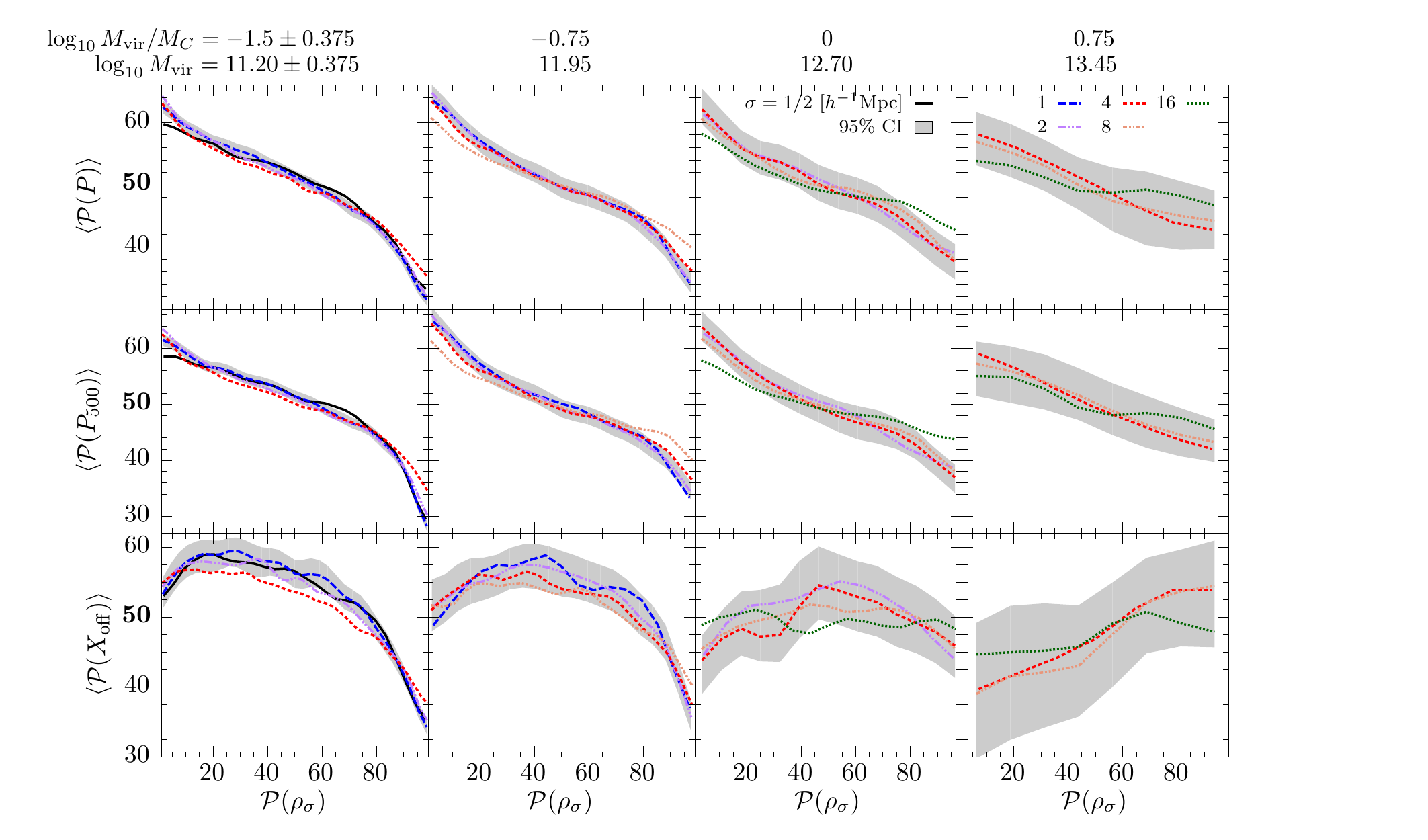}
    \caption{Same as Fig. \ref{fig:z0_correlations_p}, but showing prolateness at $\rvir$ ($P$), prolateness at $R_{500c}$ ($P_{500c}$), and the offset of halo center of mass from halo peak density ($\xoff$).  We see that prolateness measured at both $\rvir$ and $R_{500c}$  monotonically decrease with increasing local density density for all masses and all smoothing scales.  Prolateness is one of only a few halo properties (along with halo mass and Tidal Force (Fig. \ref{fig:z0_correlations_p_2} Row 3)) that exhibit a clear monotonic relationship with with local density.  That halos are more prolate at lower density may be because they form along thinner filaments at lower densities.
We find that for low mass halos $\xoff$ is lowest in high density regions and highest in median density regions, indicating that halos tend to have more mass asymmetry in median density regions and less in high density regions.  The decrease of $\xoff$ in higher density regions parallels that of prolateness, so lower mass halos are rounder and better centred at higher density, but the decrease of $\xoff$ at low densities implies that the increasingly prolate halos are also somewhat better centred at low density.  The parallel behaviour of $\xoff$ and the half-mass scale factor ($\amhalf$, Fig. \ref{fig:z0_correlations_p_2} Row 1) suggests a connection between $\xoff$ and the timing of halo formation in regions of different density.  Note also the similar behaviour as a function of density of $\xoff$ and the spin parameters $\lambdap$ and $\lambdaP$ (Fig. \ref{fig:z0_correlations_p} Row 2 and Fig. \ref{fig:z0_correlations_p_s} Row 2).
}
    \label{fig:z0_correlations_p_4}
\end{figure*}

\subsection{Spin Parameter}
\label{sec:spin}

% Recap main trends seen from density correlation plot and questions raised

	Next, we investigate the local density dependence of spin parameter.  {\color{black}We present results for both the Bullock spin parameter \citep[][]{Bullock+2001} and the Peebles spin parameter \citep{Peebles1968}, defined as
\begin{equation}
\lambdap = \frac{J}{\sqrt{2} \mvir V_{\mathrm{vir}} \rvir},
\label{eq:spinb}
\end{equation}
and
\begin{equation}
\lambdaP = \frac{J \left | E \right |^{1/2}}{G \mvir^{5/2}},
\label{eq:spinp}
\end{equation}
respectively, where $J$ and $E$ are the total halo angular momentum and energy, and $G$ is the gravitational constant.  However, we focus our analysis on $\lambdap$.}
	
	We see from Fig. \ref{fig:z0_correlations_p} Row 2 that at $z = 0$ low mass halos ($\mvir < \mstar$) in high density regions have lower spin parameters than halos in median density regions.  Similarly, halos in low density regions have distinctly lower $\lambdap$ than halos in median density regions for all masses.  We note that the downturn in spin parameter for halos in low density regions occurs at roughly the same density as the upturn in concentration for these same halos (Fig. \ref{fig:z0_correlations_p} Row 1), suggesting that the underlying cause of these trends may be related.  The relation between local density and $\lambdaP$ is very similar (Fig. \ref{fig:z0_correlations_p_3} Row 2), though with slightly higher $\lambdaP$ in high density regions compared to median density regions.  We then focus our analysis on the following questions: (1) why do lower mass halos in high density regions typically have lower spin parameters compared to halos in median density regions; and (2) why do halos in low density regions typically have lower spin parameters than halos in median density regions?

% Discuss relevant trends in density correlation with almm, (shape??)

	One of the principal mechanisms for halos to acquire angular momentum is through mergers \citep[e.g.,][]{Vitvitska+02}.  Understanding the variation in halo merger rate in regions of different density may be useful in explaining some of the dependence of spin parameter on local density.  From the correlation between scale factor of the last major merger ($\almm$) and local density, plotted in Fig. \ref{fig:z0_correlations_p_2} Row 2, we see that halos in very low density regions last experienced a major merger at earlier times than halos in median density regions.  This suggests that halos in low density regions may have lower spin parameters partly due to reduced frequency of major mergers at late times compared to halos in median density regions.  However, halos in high density regions last experienced major mergers on very similar time-scales to halos in median density regions, indicating that differences in their spin parameter distributions must be a result of other mechanisms.

% Discuss relevant trends in progenitor history plot (spin, peebles spin, tidal force, stripping (accretion/mass growth)

	We now look at the time evolution of spin parameter for halos in different density regions at $z = 0$, plotted in Fig. \ref{fig:z0_density_prog_hist_1} Row 2.  Low mass halos in high density regions at $z = 0$ have only recently developed reduced spin parameters; in fact, before $z \sim 1$ these halos had typical spin parameters equal to or greater than the progenitors of halos in median density regions at $z = 0$.  This reduction in spin parameter at late times for halos in high density regions appears coincident with the increase in concentration and reduction in accretion rate relative to halos in lower density regions, all of which likely stem from the relative increase in tidal forces around this same epoch.  On the other hand, halos in low density regions display consistently lower median spin parameters throughout their evolution than halos in higher density regions.  This indicates that the lower spin parameters in low density regions originate early in the formation history of these halos.  Halo formation in regions of lower local density may result in typically lower spin parameters due to reduced tidal torques {\color{black}\citep{Peebles1968,White1984,Porciani+2002}} compared to halos forming in higher density regions.  {\color{black}High mass halos display nearly constant median $\lambdap$ with redshift.}  We also show the time evolution of the Peebles spin parameter (Fig. \ref{fig:z0_density_prog_hist_4}), which displays similar differences between halos in different density environments, but has a different overall redshift dependence than the Bullock spin parameter \citep{Paper1}.  {\color{black}We reiterate that median trends should not be confused with individual halo evolution.  Spin parameters of individual halos can fluctuate considerably throughout their lifetimes due to accretion events and stripping.}

% Discuss stripped halos involvement (in density correlation and stripping vs spin)

	In order to test the effect that halo stripping has on spin parameter, we plot in Fig. \ref{fig:z0_correlations_p_ns} the correlation between spin parameter and local density, excluding halos that have lost more than 2\% of their peak mass.  While medians of spin parameter are determined using only un-stripped halos, we compute percentiles relative to all halos in order to make a fair comparison to the all-halo correlation (Fig. \ref{fig:z0_correlations_p} Row 2). We see that in high density regions where stripping is most common, halos that are not appreciably stripped have higher spin parameters.  This effect is most prominent for low mass halos, since higher mass halos are less likely to be strongly stripped.  Thus, the downturn in spin parameter for low mass halos in high density regions is strongly correlated with these halos being stripped, a process that is much less frequent in lower density regions.

% Recap on what we know, what questions we still have

	We've seen that many halos in high density regions evolved from regions of lower density.  The transition into very high density and high tidal force regions results in suppression of accretion rates and eventual stripping of material from the outer regions of many halos.  Suppressed accretion rates are accompanied by decreased spin parameters, possibly as a result of halos preferentially losing eccentric, high angular momentum particles.  Halos in low density regions have occupied low density regions for most of their evolution.  We suggest that the underdense regions in which these halos formed exerted weaker tidal torques on collapsing protohalos, which effectively set the typical spin parameter of these halos.  Spin parameter has high dispersion in all environments (Fig. \ref{fig:z0_correlations_s}), and can vary widely over the lifetime of an individual halo due to sensitivity to accretion events such as major mergers.  Nevertheless, low tidal torques exerted on halos throughout their lifetimes seems to be a plausible explanation for the lower spin parameters of low mass halos in low density regions at $z = 0$.

\begin{figure*}
	\centering
	\includegraphics[trim=18 180 60 160, clip, width=\textwidth]{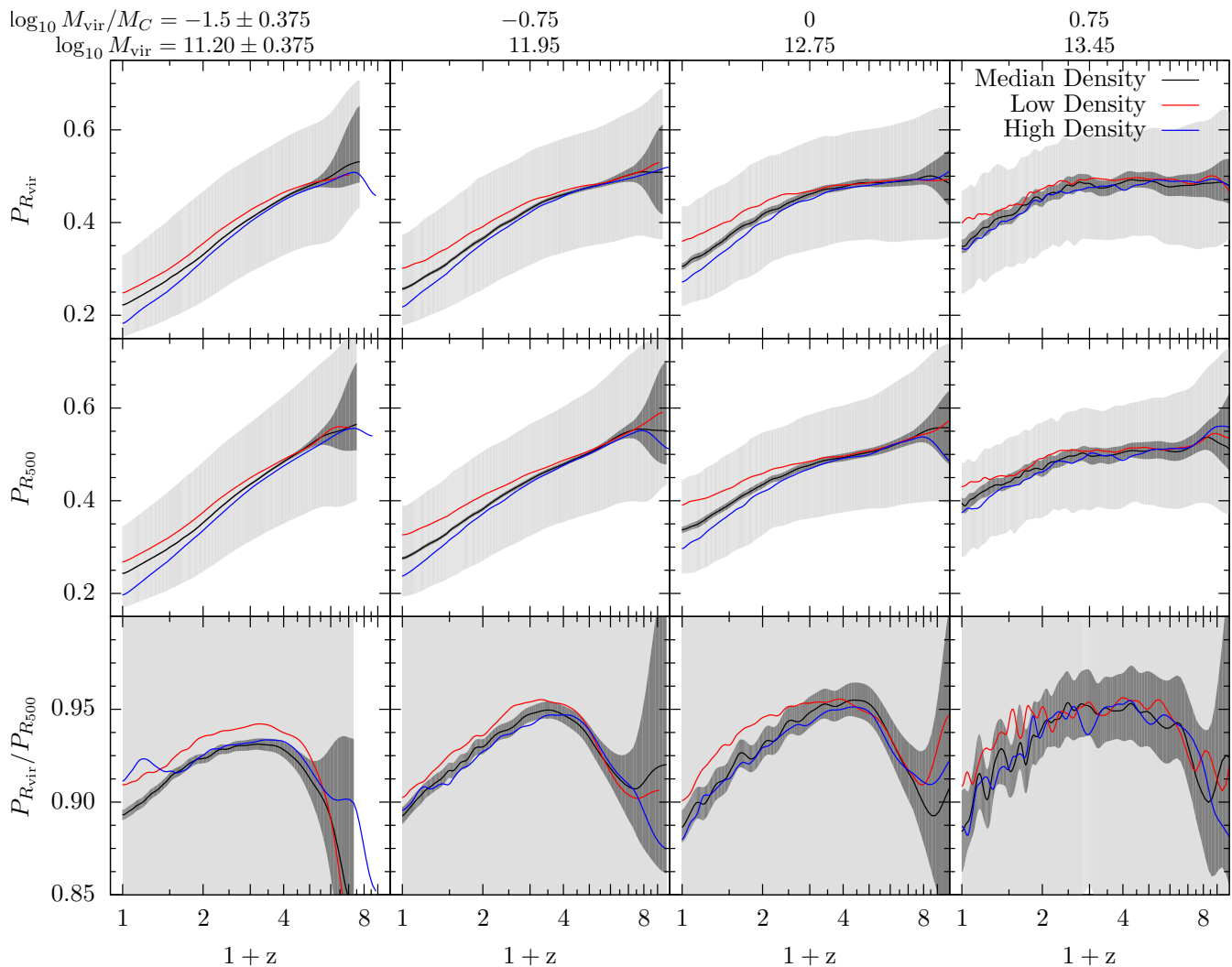}
    \caption{Same as Fig. \ref{fig:z0_density_prog_hist_1}, but showing prolateness measured at $\rvir$ ($\pvir$), prolateness measured at $R_{500c}$ ($P_{R_{500c}}$), and the ratio of these two.  We see that all halos are more prolate at high redshift, and become less prolate over time.  Halos in high density regions at $z = 0$ tend to sphericalize more quickly than halos in low density regions, which tend to sphericalize the slowest.  This is true for prolateness measured at both $\rvir$ and $R_{500c}$.  The ratio $\pvir / P_{R_{500c}}$ tells us that since $z \lesssim 3$, the outer regions of halos become rounder more quickly than the inner regions.  Additionally, lower mass halos in low density regions have the least change in shape from the inner ($P_{R_{500c}}$) to outer ($\pvir$) regions since early in their formation history, while halos in median and high density regions at $z = 0$ show a greater disparity between inner and outer shape throughout most of their evolution.}
    \label{fig:z0_density_prog_hist_3}
\end{figure*}

\subsection{Prolateness}
\label{sec:shape}

% define prolateness
Finally, we take a look at how halo shape varies with environment density.  We've defined a new halo shape parameter, \textit{Prolateness}, defined as
\begin{equation}
P \equiv 1-\frac{1}{\sqrt[]{2}}\left[\left(\frac{b}{a}\right)^{2} + \left(\frac{c}{a}\right)^{2}\right]^{1/2},
\label{eq:prolateness}
\end{equation}
such that $1-P$ is the magnitude of the vector $(\frac{b}{a},\frac{c}{a})$ normalized by $\frac{1}{\sqrt{2}}$, where $a$, $b$, and $c$ are the lengths of the largest, second largest, and smallest triaxial ellipsoid axes, respectively, determined using the weighted inertia tensor method of \citet{Allgood+2006}.  Prolateness ranges from 0 (perfect sphere) to 1 (maximally elongated, i.e. a needle), with most halos falling somewhere in the range of $0.2 - 0.6$.

% explain trends in density - shape relation and questions raised

We see from Fig. \ref{fig:z0_correlations_p_4} Rows 1 and 2 that prolateness measured at both $\rvir$ and $R_{500c}$ (where $R_{500c}/\rvir$ is typically about 0.7) decreases monotonically with increasing density.  Halos in low density regions are more prolate and halos in high density regions are less prolate.  Interestingly, this is true for all of the mass bins we present, suggesting that a universal phenomenon underlies this relationship.  We attribute the rounder halos in high density regions partly to stripping by massive neighbours.  Tidal stripping would tend to preferentially remove loosely bound particles with more elliptical orbits, resulting in rounder halos.  However, we would expect this to more dramatically affect $\pvir$ than $P_{R_{500c}}$, which does not seem to be the case.  $P_{R_{500c}}$ is equally, if not slightly more correlated with density than $\pvir$, especially in high density regions.  This suggests a more general phenomenon may govern the evolution of halo prolateness, such as accretion geometry.  Total accretion rate (Fig. \ref{fig:z0_correlations_p} Row 3) displays a similar relationship as prolateness from median to high density regions, but is largely flat for halos in regions below median density.  We expect then, that prolate halos in low density regions differ from rounder halos in higher density regions by the nature of how they accrete material, but not their total accretion rate.  For example, halos in low density regions may accrete a larger fraction of material along a preferred axis, or reside in thinner filaments, compared to halos in median density regions, which would tend to build prolate halos \citep[e.g.,][]{Allgood+2006,Vera-Ciro+11,Despali+14,Despali+16}.  Halos in high density regions may be rounder due to a combination of more isotropic accretion of material, and tidal stripping of material with highly eccentric orbits.  Altogether, these trends pose one primary question for clarification: is there a single underlying mechanism that drives the relationship between local density and halo prolateness?

% explain trends in shape evolution plot and questions answered / raised
Looking at the evolution of halo prolateness for halos that end up in different density regions at $z = 0$ (Fig. \ref{fig:z0_density_prog_hist_3}), we see that all halos grow less prolate with time, but those in higher density regions do so more rapidly than those in lower density regions.  This is true for prolateness measured at both $\rvir$ and $R_{500c}$, and for $\mvir \lesssim \mstar$.  We note that shapes of halos in high density regions at $z = 0$ begin to diverge from those in median density regions around redshift ($z \sim  1.5$), which is roughly when these halos begin to experience dramatically higher tidal forces (see Fig. \ref{fig:z0_density_prog_hist_2} Row 1).  Low mass halos in low density regions begin to sphericalize at a slower rate than halos in median density regions early in their evolution ($z \sim 2.5$).  Overall, we observe a gradual divergence in sphericalization rate between halos in high density regions and those in low density regions, starting around $z \sim 1.5-2.5 $ for halos less massive than the critical mass $\mstar$.  This contrasts with certain other halo properties, such as $\cnfw$ and $\lambdap$, where we see sharp deviations at late times for halos that end up in high density regions.  Since we expect tidal stripping to be chiefly responsible for these sharp trends in $\cnfw$ and $\lambdap$, the absence of dramatic differences in prolateness at late times suggests tidal stripping may not be the main underlying reason for halos rounding out most rapidly in high density regions.

In Fig. \ref{fig:z0_density_prog_hist_3} Row 3, we plot the evolution of the ratio of prolateness measured at $\rvir$ to prolateness measured at $R_{500c}$.  This gives us a sense for how the shape of the inner region of the halo is changing relative to the outer region.  The ratio $\rvir/R_{500c}$ is less than unity; it is well known \citep[e.g.,][]{Allgood+2006} that halos are somewhat more prolate at smaller radii.  In general, we see that the outer regions of halos tend to sphericalize more rapidly than the inner regions (at least, for $z \lesssim 4$).  For low mass halos, we see that those in low density regions have less difference between their inner and outer shapes for most of their history compared to those in higher density regions.   
%Our interpretation of 
The slight bump that halos moving into high density regions experience at late times ($z \sim 0.25$) indicates a rapid "inside-out" rounding -- initially their inner regions, followed by subsequent rounding of the outer regions.  We expect this phenomenon to be related to harassment by more massive halos and tidal stripping.

% Recap on what we know, what questions remain
Overall, we find that halos in high density regions at $z = 0$ are rounder than those in low density regions, but have a greater difference between their inner (more prolate) and outer (less prolate) shapes.  We expect that the nature of the accretion rate of halos (including where they are located in the cosmic web) underlies these differences, but that tidal stripping from massive halos in high density regions also plays a part in modulating halo shape.

%%%%%%%%%%%%%%%%%%%%%%%%%%%%%%%%%%%%%%%%%%%%%%%%%%%%%%
\section{Discussion and Conclusions}

We investigate how properties of distinct dark matter halos in the Bolshoi-Planck simulation depend on the density of their surrounding environment.  We determine local densities using a CIC method of counting particles on a $1024^3$ grid and then smoothing with a Gaussian kernel with several different smoothing radii.  In \S\ref{sec:Density_Distributions}, we plot distributions of cosmic densities and find that they are well fit by a Generalized Extreme Value Distribution.  We find that at smaller smoothing scales, the density distributions peak at lower densities and sample more non-linear structure (voids) than larger smoothing scales, which peak closer to the average density of the simulation and span a narrower total range of densities.

In \S\ref{sec:HMF_density}, we present halo mass functions in different density regions for $z = 0$, 0.5, 1, and 2.  We find that mass functions in lower density regions have lower characteristic masses than in higher density regions.  The mass functions extend to higher masses and have higher characteristic masses at low redshift, especially in high density regions, but otherwise evolve little from z = 2 to z = 0.

\S5 begins by summarizing correlations between various halo properties and the environmental density around the halo, and the redshift evolution of halo properties at different densities.
In \S\ref{sec:mar} through \S\ref{sec:shape}, we present results on how halo mass accretion rate ($\dot{M}_{\tau_{\mathrm{dyn}}}/M$), NFW concentration ($\cnfw$), spin parameter ($\lambdap$), and shape ($P$) depend on the local density smoothed on several length scales around the halos.  For each halo mass range we choose a characteristic smoothing scale that is at least 8 times greater than the typical virial radii of the halos, allowing us to capture local effects that are lost when smoothing on larger scales, while not being too influenced by the halos' own density profiles.  In Fig. \ref{fig:z0_correlations} we plot the relationship between local density and $\cnfw$, $\lambdap$, and specific accretion rate, which we plot again but in percentilized form in Fig. \ref{fig:z0_correlations_p}.  We find that low mass halos ($\mvir < \mstar$) in high density regions accrete significantly less than halos in lower density regions.  The median accretion rate in the highest density regions even drops below zero, indicating that a majority of low mass halos in very high density regions are losing mass due to being tidally stripped by nearby massive neighbours.  Halos in low density regions accrete at similar rates as those in median density regions.  Higher mass halos have a less dramatic drop in accretion rate in high density regions for two reasons:  
\begin{enumerate}
\item The highest mass halos are the ones stripping the lower mass halos, and so should have high accretion rates, unless they are in the vicinity of an even more massive halo.  Also, massive halos tend to be physically separated from other massive halos, making them less likely to be stripped.
\item High mass halos occupy exclusively high density regions.  Any trends observed will then only span a narrow range of densities, and be limited in dynamic range as a result. 
\end{enumerate} 
These points (lack of stripping and narrow range of mostly higher than average densities) account for the main differences between halos below $\mstar$ and above $\mstar$ for most of the halo property--density relations we observe.  

\citet{Maulbetsch+07} found that halos of the same final mass accreted their mass earlier in denser environments, and also accreted a significantly higher fraction of their mass in major mergers. We have not investigated the fraction of mass accreted in major mergers in this work, but do find that halos accreted their mass earlier in high density regions.  Since halos in high density regions have suppressed accretion rates at $z = 0$, their progenitors must be more massive than the progenitors of halos in lower density regions, and thus will have formed earlier on average.  Since the total accretion rate of halos in low density regions is similar to halos in median density regions, we expect that the fraction of mass accreted in mergers may play a role in differentiating halo properties in low and median density regions.

In \S\ref{sec:cnfw}, we find that low mass halos in high density regions have dramatically higher NFW concentrations than halos in lower density regions.  This trend has been well established in the literature \citep[e.g.,][]{Bullock+2001,Avila-Reese+05,Maccio+07}, and is primarily caused by tidal stripping.  This tends to remove material from the outer regions of halos, resulting in steeper outer profiles, poorer NFW fits, and hence artificially reduced scale radii and raised NFW concentrations.  We find that the minimum median concentration varies with local density, and is in lower density regions for lower mass halos.  In the lowest mass bin ($\mvir = 10^{11.2} \msun$), halos in the lowest density regions have concentrations about $30\%$ higher (about $10\%$ difference in percentile) than the minimum median concentration, which is around the 20th percentile in local density.  We find a corresponding decrease in half-mass scale factor (earlier formation time) in low density regions, and also see that halos in low density regions have had a consistently elevated median concentration compared to halos in median density regions.  This suggests that the halos in the lowest density regions formed earlier on average and have had more time to grow their concentrations than halos in somewhat higher density regions.

We find that the spin parameter (\S\ref{sec:spin}) is about $30\%$ lower in the lowest density regions than in median density regions for all halo masses.  Furthermore, halos in the lowest density regions have had consistently lower spins (Fig. \ref{fig:z0_density_prog_hist_1}), and experienced consistently lower tidal forces (Fig. \ref{fig:z0_density_prog_hist_2}) throughout their evolution.  The magnitude of this effect is similar to that of the increase in concentration in the lowest density regions, and so may be related to differences in the nature of the accretion histories of halos in different density regions (amount of material accreted in lumps, etc.).  The systematic offset in median spin parameter reflects lower tidal torques exerted on halos in low density regions due to forming in underdense regions.  We therefore expect that low mass halos with lower spins in low density regions would tend to host rotationally supported galaxies with less extended disks, and possibly also smaller early-type galaxies as well \citep[e.g.,][]{Kravtsov2013}.  This could be observationally tested by comparing galaxy sizes and morphologies at the same stellar mass in low and median density regions.

In high density regions, we find that low mass halos have reduced spin parameters and high mass halos have increased spin parameters compared to halos in median density regions.  Low mass halos tend to be heavily stripped in high density regions at $z\sim0$ (Fig. \ref{fig:z0_correlations}).  We expect that tidal stripping preferentially removes the high angular momentum particles from the outer regions of halos, reducing their spin parameters.  This role of tidal stripping in suppressing spin parameter is supported in Figs \ref{fig:z0_correlations_p_ns} and \ref{fig:z0_correlations_p_s}, where we see that low mass halos that have not been appreciably stripped do not have lower spin parameters in high density regions than in median density regions.   Because this stripping occurs mainly at low redshift, after most galaxy formation has ended in central galaxies in low mass halos in dense regions, we do not expect that it will affect the sizes of these galaxies.  \citet{WangMoJingYangWang11} found that halos in higher tidal field environments and in higher density environments (smoothed in spheres of radius $6 \ \hmpc$) have higher spins, with the trends stronger for higher mass halos.  This is consistent with our results for halo masses $ \grtsim \mstar$.  We would expect to see some downturn in spin in high density regions for their lowest mass bin ($\mvir = 10^{12}-10^{12.5} \ \msun$), but this would be a quite small effect and is likely within the statistical error of the analysis.  We see stronger suppression of spin in high density regions for lower masses (and smaller smoothing scales) than studied by \citet{WangMoJingYangWang11}.

We find in \S\ref{sec:shape} that halo prolateness is monotonically dependent on density.  Halos in the lowest density regions are typically the most elongated at both $\rvir$ and $R_{500c}$, become rounder at the slowest rate, and have the least change between inner ($R_{500c}$) and outer ($\rvir$) shape throughout their evolution.  Halos in the highest density regions are the roundest at both radii measured and sphericalize the most rapidly, but have a relative change between inner and outer shape consistent with halos in median density regions.  We also see that more massive halos tend to be more elongated at a given redshift, and have less relative change between inner and outer shape than lower mass halos.  These results are largely consistent with other works that have studied how halo shape varies with radius (or overdensity within the halo) and evolves over time \citep[e.g.,][]{Allgood+2006,Vera-Ciro+11,Despali+14,Despali+16}, which find that the shape of a shell at a given overdensity reflects the nature of the accretion at the epoch when that shell primarily assembled.  {\color{black}Early in their evolution, halos tend to build elongated shells due to highly directional accretion along narrow filaments in lower density regions.  As their host filaments grow thicker, they experience more isotropic merging and overall suppressed accretion of fast moving material and tend to build more spherical shells in higher density regions \citep[see, e.g.][]{B+16}.  In future work, we intend to further investigate how halo properties correlate with their location in the cosmic web.}

The results we present in this paper are consistent with \citet{Avila-Reese+05}, which found that halos in cluster environments had lower spin parameters, higher concentrations, and are more round than halos in the field.  \citet{Maccio+07} found that higher-concentration low-mass halos are found in denser environments, and lower-concentration ones in less dense environments, which helps to explain assembly bias, {\color{black}i.e. that higher-concentration early-accreting halos with mass $\mvir < \mstar$ are more clustered \citep{GaoSpringelWhite05,GaoWhite07,Wechsler+2006}.}  However, they found only a weak dependence of mean axis ratio on environment (with high density regions having slightly rounder halos), and no significant difference in spin parameter in different density regions.  This is not inconsistent with our results, but highlights the necessity for better statistics to adequately constrain trends in regions of extreme high and low density, especially for high dispersion halo properties like spin parameter.

%%%%%%%%%%%%%%%%%%%%%%%%%%%%%%%%%%%%%%%%%%%%%%%%%%%%%%
\section*{Acknowledgements} We acknowledge stimulating discussions with Vladimir Avila-Reese, Sandra Faber, and David Koo, and we thank the referee for helpful comments and questions.
CTL and JRP were supported by CANDELS grant HST GO-12060.12-A, provided by NASA through a grant from the Space Telescope Science Institute (STScI), which is operated by the Association of Universities for Research in Astronomy,  Incorporated, under NASA contract NAS5-26555.  PB was partially supported by a Giacconi Fellowship from STScI.  The remainder of support for PB through program number HST-HF2-51353.001-A was provided by NASA through a Hubble Fellowship grant from STScI. ARP was supported by UC-MEXUS Fellowship. AD was supported by NSF grants AST-1010033 and AST-1405962. Computational resources supporting this work were provided by the NASA High-End Computing (HEC) Program through the NASA Advanced Supercomputing (NAS) Division at Ames Research Center, and by the Hyades astrocomputer system at UCSC.  

%%%%%%%%%%%%%%%%%%%%%%%%%%%%%%%%%%%%%%%%%%%%%%%%%%
%%%%%%%%%%%%%%%%%    REFERENCES    %%%%%%%%%%%%%%%%%%

\bibliographystyle{mn2efix.bst}
\bibliography{blib,Bolshoi-Planck,density}


%\bibliographystyle{mnras}

%\begin{thebibliography}{99}
%\end{thebibliography}

%%%%%%%%%%%%%%%%%%%%%%%%%%%%%%%%%%%%%%%%%%%%%%%%%%

%%%%%%%%%%%%%%%%% APPENDICES %%%%%%%%%%%%%%%%%%%%%

\appendix
\section{}
\begin{figure*}
	\includegraphics[trim=5 20 25 25, clip, width=\columnwidth]{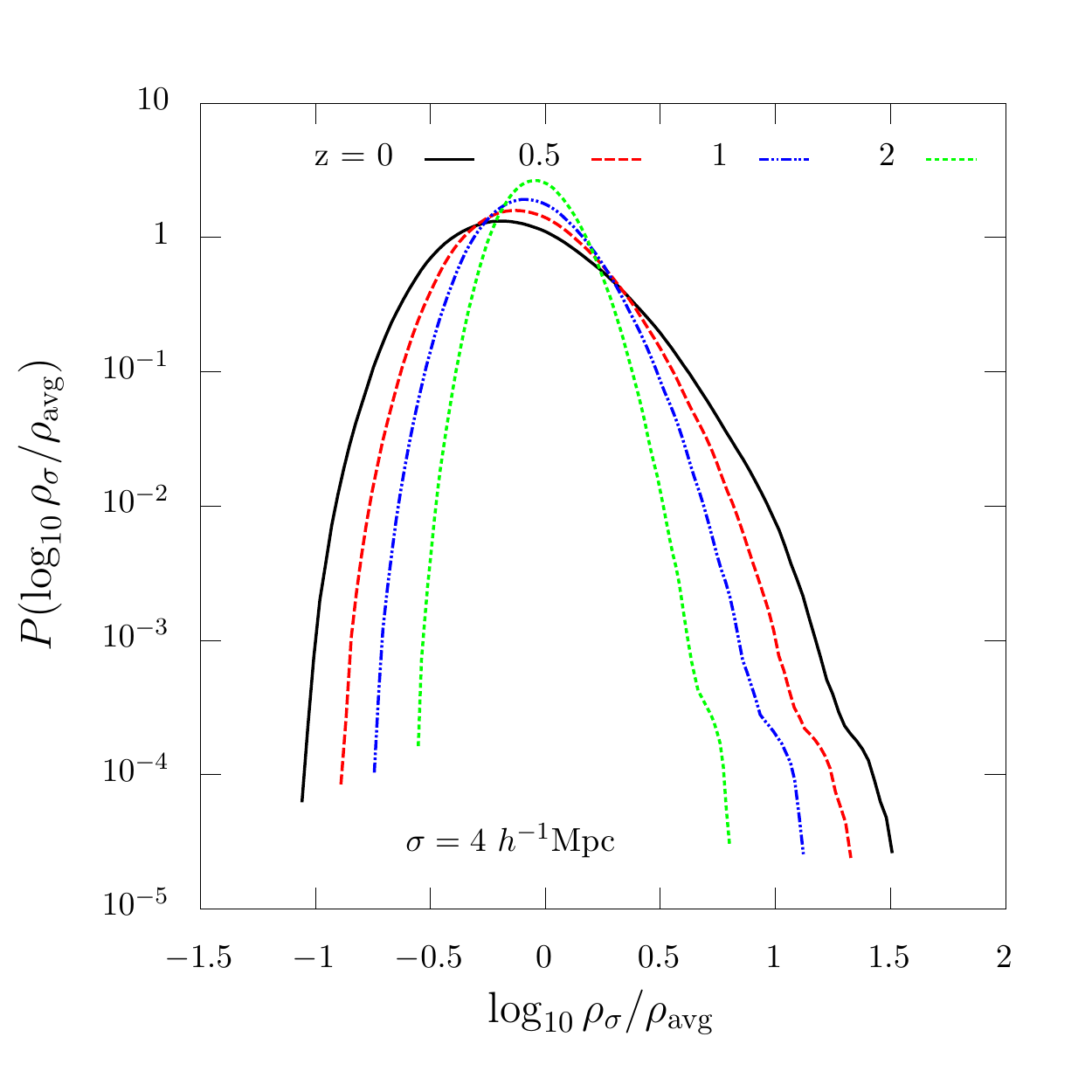}
    \caption{Probability distributions of local environment density smoothed using $\sigma_{s} = 4 \ \hmpc$ for the entire simulation volume, shown with log scaling on the vertical axis. Different coloured lines represent the same smoothing scale, but at different redshifts.  Voids grow emptier with time, shifting the peak to lower densities.  Non-linear structure grows as redshift decreases, but not as dramatically as on smaller scales (Fig. \ref{fig:density_distr_log_z}).}
    \label{fig:density_distr_log_z_sigma4}
\end{figure*}

\begin{figure*}
	\centering
	\includegraphics[trim=0 12 10 12, clip, width=\columnwidth]{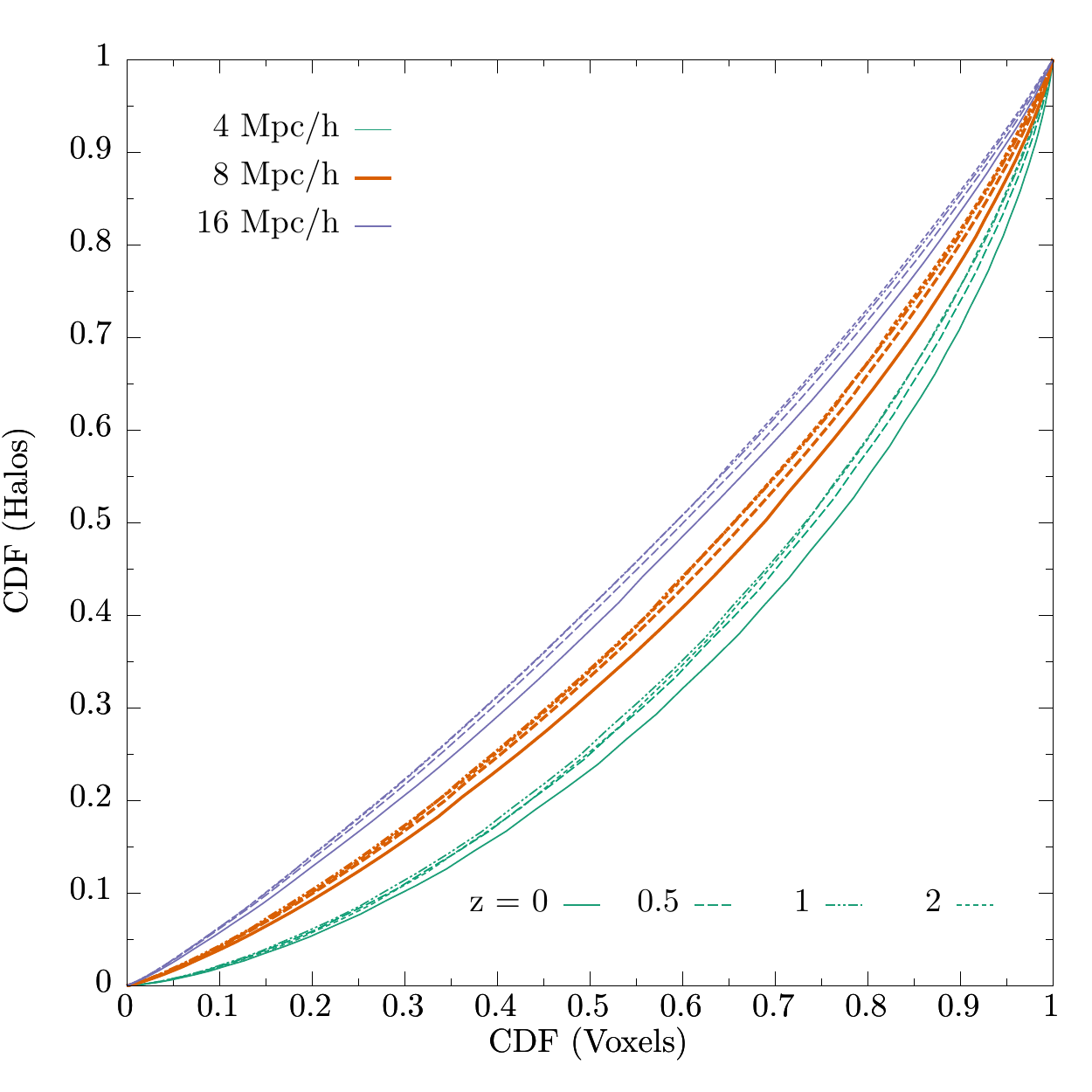}
    \caption{Association between percentilized cosmic local densities and halo local densities, smoothed on scales $\sigma = 4$, 8, and $16 \ \hmpc$, and for redshifts 0, 0.5, 1, and 2.  Cumulative distribution functions indicate how percentiles of density in the full simulation volume translate to percentiles of density around halos ($\mvir > 10^{10} \ \msun$).  We see that the distributions of halo local densities and full volume densities are most similar at $z = 2$, and least similar at $z =0$.  Halos tend to accumulate in higher percentiles of full volume density at low redshift compared to at high redshift.}
    \label{fig:density_cdf_cdf}
\end{figure*}

\begin{figure*}
	\centering
	\includegraphics[trim=20 55 35 65, clip, width=\textwidth]{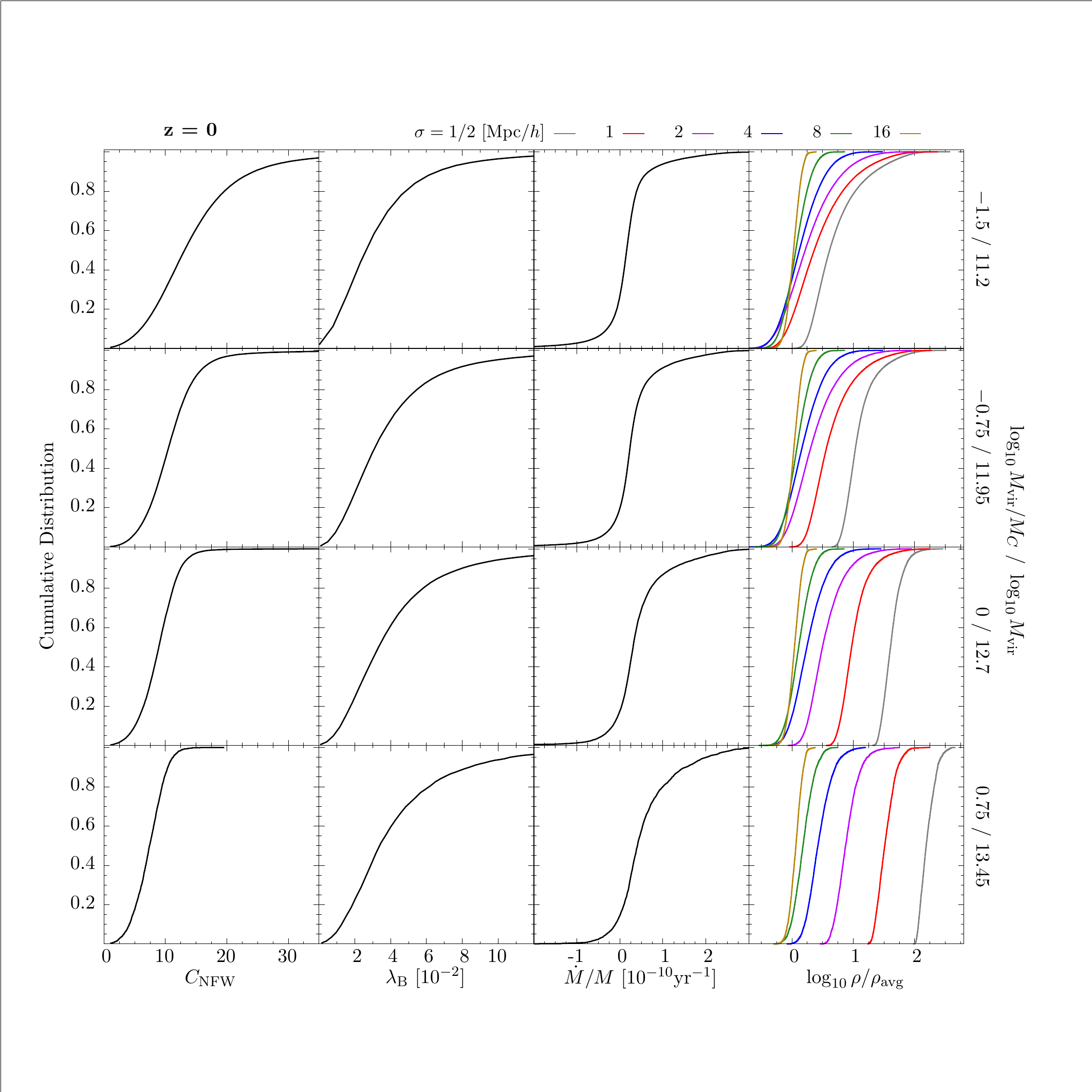}
    \caption{Cumulative distribution functions of NFW concentration, spin parameter, specific accretion rate, and smoothed local density parameter, shown at z = 0 with the same mass bins as Fig. \ref{fig:z0_correlations_p}.  Each row corresponds to an individual mass bin, with mass labels indicated on the right vertical axis. This figure relates the percentilized halo property--density correlation plot (Fig. \ref{fig:z0_correlations_p}) to the real valued halo property--density correlation plot (Fig. \ref{fig:z0_correlations}).}
    \label{fig:z0_density_cdf}
\end{figure*}

\begin{figure*}
	\centering
	\includegraphics[trim=21 10 85 10, clip, width=\textwidth]{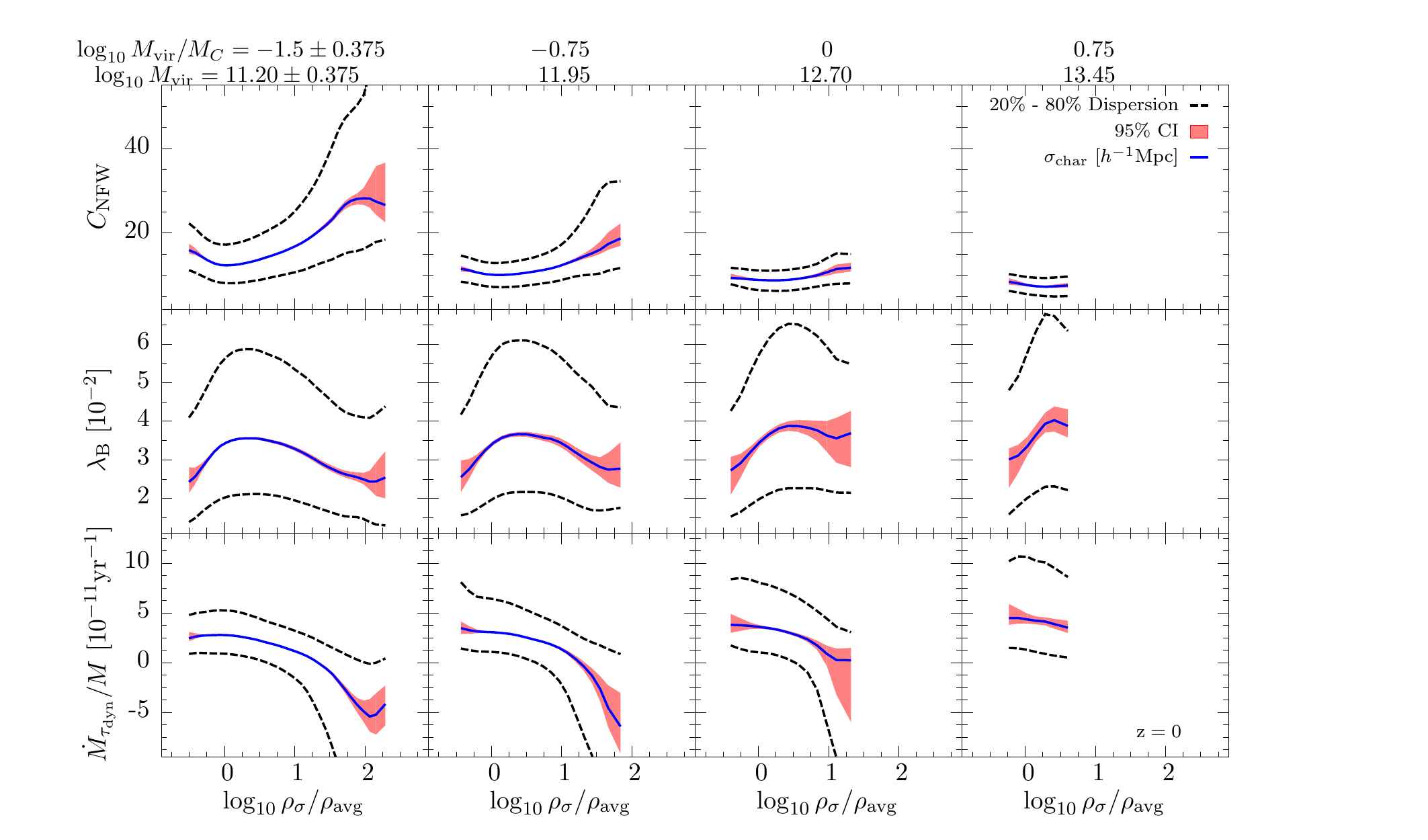}
    \caption{Same as Fig. \ref{fig:z0_correlations_p}, but with dotted black lines showing 20-80 percentile range scatter in $\cnfw$, $\lambdap$, and $\dot{M}/M$.  The pink shaded area indicates the 95\% confidence interval on the median.  Only the 1/2 $\hmpc$ smoothed density parameter is shown.  The scatter tends to be greater at high densities than low densities for $\cnfw$ and accretion rate, but independent of density for spin parameter.}
    \label{fig:z0_correlations_s}
\end{figure*}

\begin{figure*}
	\centering
	\includegraphics[trim=21 10 85 10, clip, width=\textwidth]{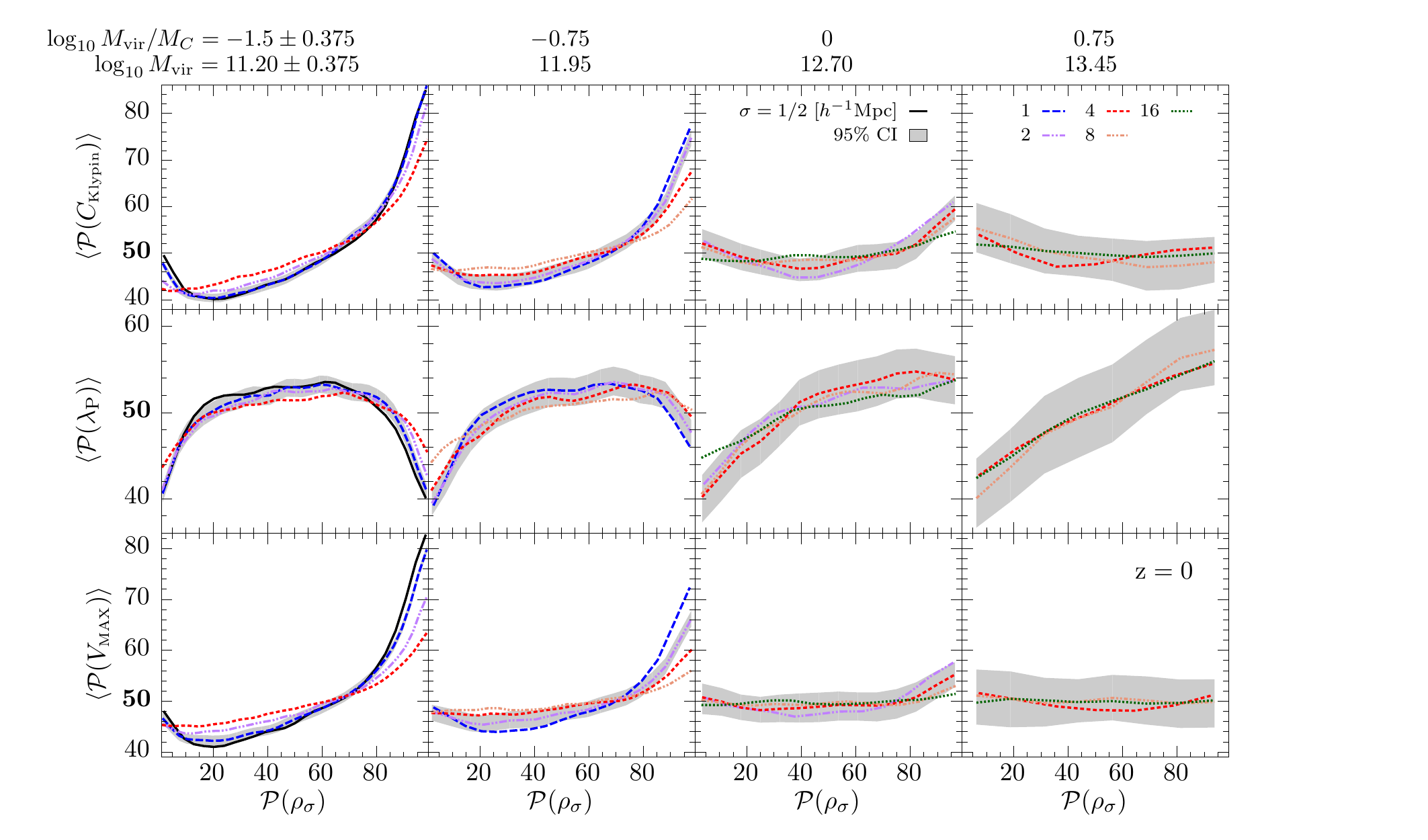}
    \caption{Same as Fig. \ref{fig:z0_correlations_p}, but showing Klypin concentration ($\cklypin$), Peebles spin parameter ($\lambdaP$), and maximum circular velocity ($\vmax$).  The $\cklypin-\rho$ relation is very similar to the $\cnfw-\rho$ relation, but with slightly less $\cklypin$ increase in high density regions.  We also see a similar relation between $\vmax$ and local density:  low mass halos in high density regions have much higher $\vmax$ than halos in lower density regions, with the reverse observed in very low density regions.  $\lambdaP$ is less reduced in high density regions and slightly more reduced in low density regions compared to $\lambdap$.}
    \label{fig:z0_correlations_p_3}
\end{figure*}

\begin{figure*}
	\centering
	\includegraphics[trim=18 150 60 160, clip, width=\textwidth]{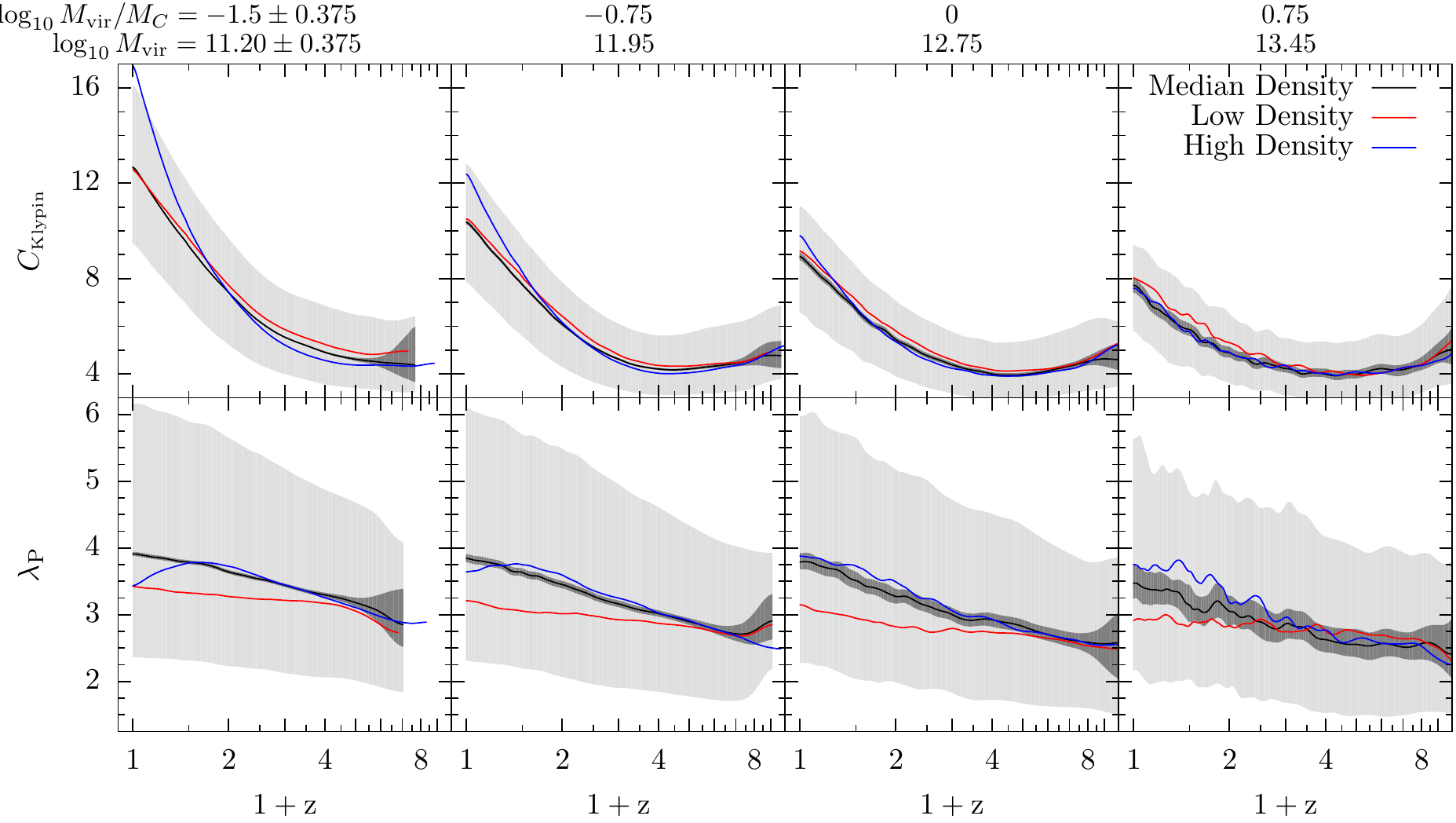}
        \caption{Same as Fig. \ref{fig:z0_density_prog_hist_1}, but showing concentration determined using the Klypin method and Peebles' spin parameter.  We observe similar trends as in \ref{fig:z0_density_prog_hist_1} Rows 1 and 2, except that median $\cklypin$ values of low mass halos in high density regions are lower than $\cnfw$, and $\lambda$ evolution tends to increase with time (tendency towards positive slope rather than negative like $\lambdap$).}
    \label{fig:z0_density_prog_hist_4}
\end{figure*}

\begin{figure*}
	\centering
	\includegraphics[trim=21 10 85 10, clip, width=\textwidth]{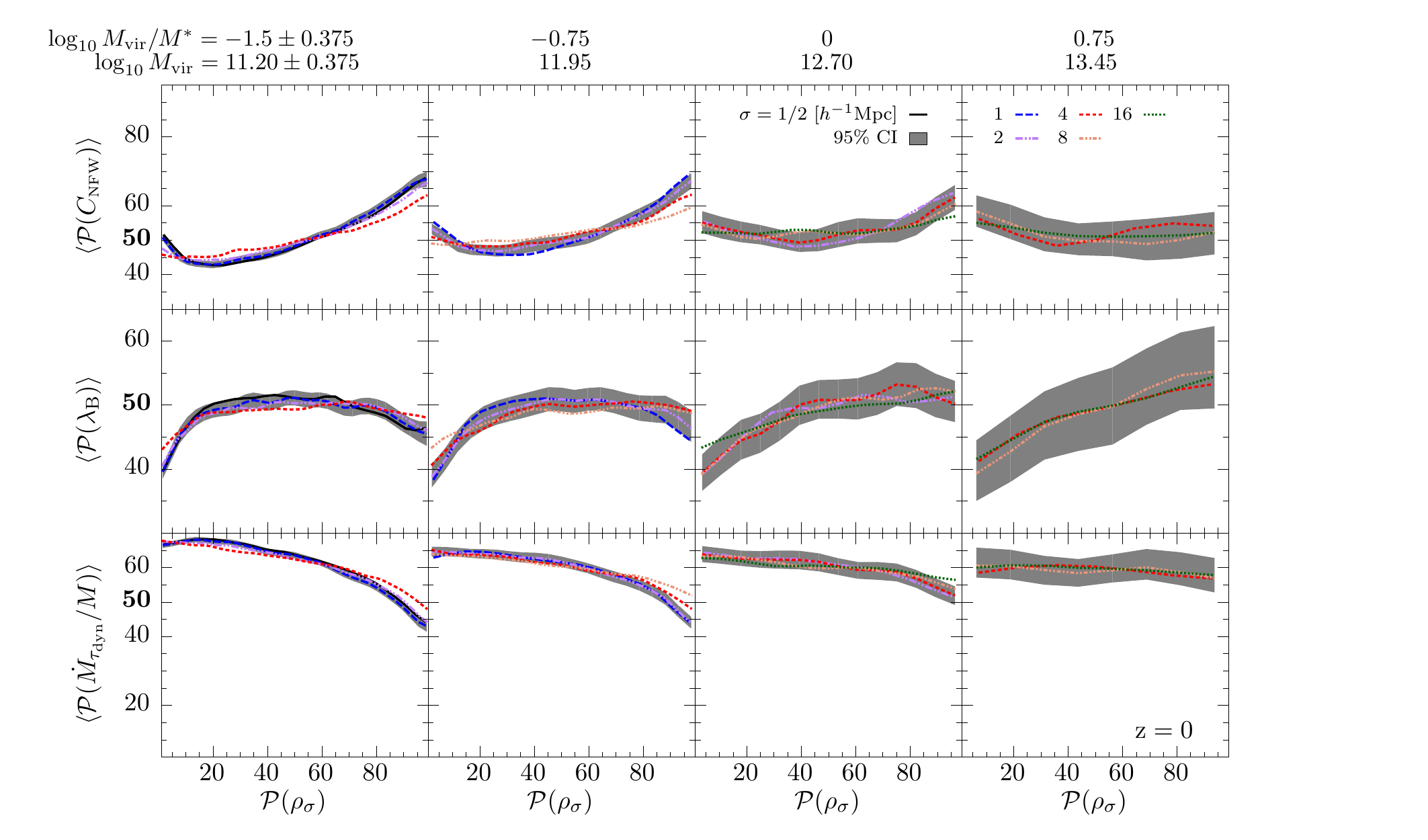}
    \caption{Same as Fig. \ref{fig:z0_correlations_p}, but does not include halos that have lost more than 2\% of their mass ($\mvir/\mpeak < 0.98$).  We determine medians using this sub-population, but we determine percentiles relative to all halos in the mass bin, allowing a fair comparison to Figs \ref{fig:z0_correlations_p} and \ref{fig:z0_correlations_p_s}.  The correlations presented are only appreciably different for low mass halos in high density regions, where we see that concentrations are lower, spin parameters are higher, and accretion rates are higher compared to the all-halo correlations.}
    \label{fig:z0_correlations_p_ns}
\end{figure*}

\begin{figure*}
	\centering
	\includegraphics[trim=21 10 85 10, clip, width=\textwidth]{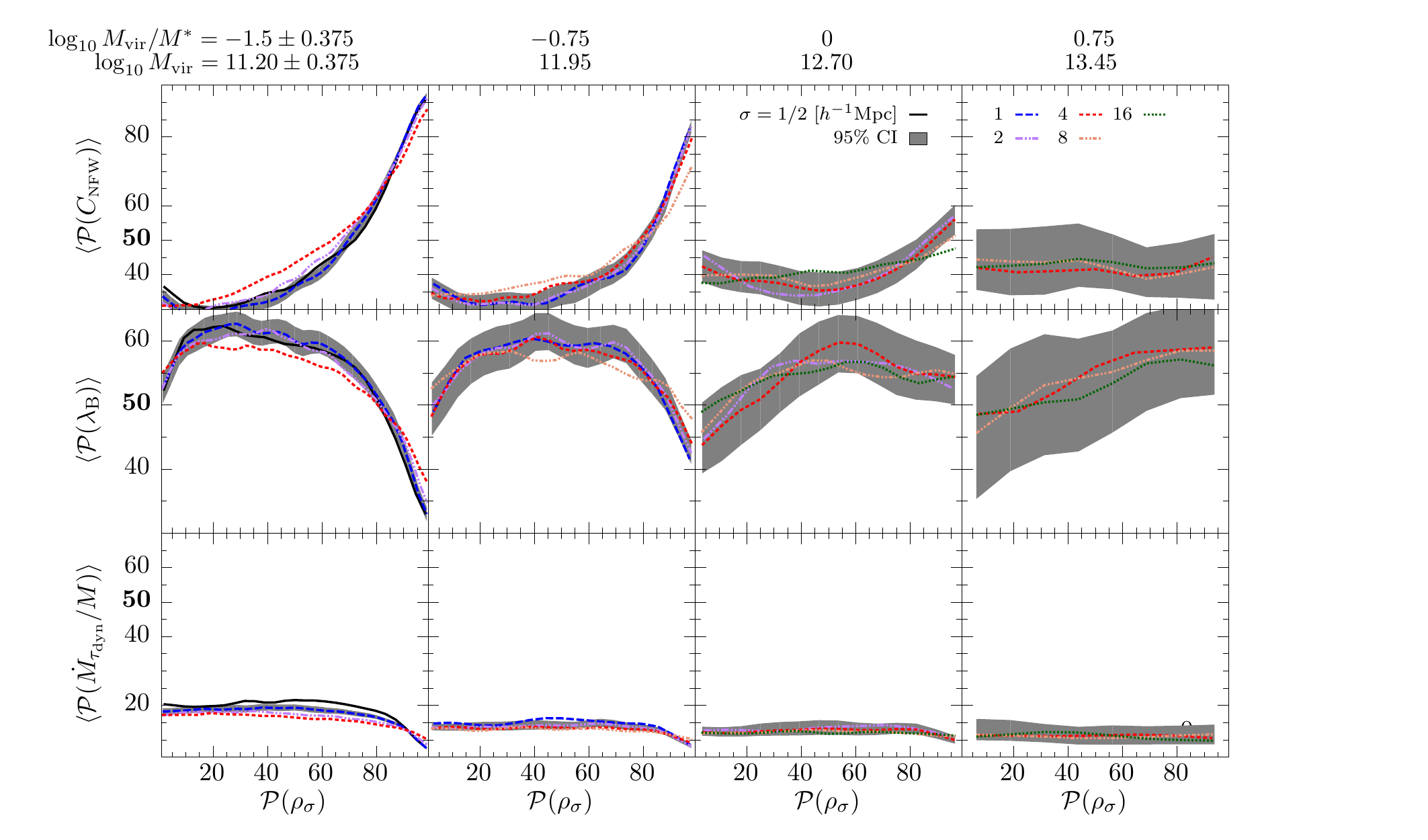}
    \caption{Same as Fig. \ref{fig:z0_correlations_p}, but only includes halos that have lost more than 2\% of their mass ($\mvir/\mpeak < 0.98$).  We determine medians using this sub-population, but we determine percentiles relative to all halos in the mass bin, allowing a fair comparison to Figs \ref{fig:z0_correlations_p} and \ref{fig:z0_correlations_p_ns}.  We see that in high density regions, low mass stripped halos have high concentrations, low spin parameters, and very low accretion rates.  In low density regions, low mass stripped halos also have low accretion rates, but lower concentrations, and higher spin parameters.  Stripping has an opposite effect on concentration and spin parameter in high density regions compared to low density regions.  Note that there are few stripped halos in low density regions compared to in high density regions.}
    \label{fig:z0_correlations_p_s}
\end{figure*}

This Appendix contains figures that supplement those in the text.  Like Fig. \ref{fig:density_distr_log_z}, Fig. \ref{fig:density_distr_log_z_sigma4} shows the probability distribution of local environment density, but smoothed on $\sigma_{s} = 4 \hmpc$ rather than $1 \hmpc$.  Fig. \ref{fig:density_cdf_cdf} illustrates the connection between percentilized cosmic local densities and halo local densities at various redshifts and smoothing scales, and Fig.  \ref{fig:z0_density_cdf} relates the percentilized halo property--density correlation plot (Fig. \ref{fig:z0_correlations_p}) to the non-percentilized halo property--density correlation plot (Fig. \ref{fig:z0_correlations}).  Fig. \ref{fig:z0_correlations_s} is a supplement to Fig. \ref{fig:z0_correlations_p}, showing the 20-80 percentile range and the 95\% confidence interval on the median at the 1/2 $\hmpc$ smoothing.  
Supplementing Figs. \ref{fig:z0_correlations_p} and \ref{fig:z0_density_prog_hist_1}, which show the NFW concentration and $\lambdap$ distribution with density at $z=0$ and their redshift evolution, Figs. \ref{fig:z0_correlations_p_3} and \ref{fig:z0_density_prog_hist_4} show the similar behaviour of Klypin concentration and $\lambdaP$.  Fig, \ref{fig:z0_correlations_p_3} also shows the distribution with density of halo maximum circular velocity $\vmax$.  Finally, Figs. \ref{fig:z0_correlations_p_ns} and \ref{fig:z0_correlations_p_s} supplement Fig. \ref{fig:z0_correlations_p} by showing only halos that have lost less than 2\% of their mass versus those that have lost more than 2\% of their mass, showing the strong effects of stripping on halo concentration, spin, and mass accretion rate, especially for lower mass halos in high density regions.

%%%%%%%%%%%%%%%%%%%%%%%%%%%%%%%%%%%%%%%%%%%%%%%%%%

% Don't change these lines
\bsp	% typesetting comment
\label{lastpage}
\end{document}